\documentclass[pra,aps,letterpaper,groupedaddress,twocolumn,superscriptaddress,floatfix]{revtex4-1}

\usepackage{mathptmx}
\usepackage[utf8]{inputenc}
\usepackage[T1]{fontenc}

\usepackage{amssymb}
\usepackage{amsmath}
\usepackage{mathtools}
\usepackage{mathrsfs}
\usepackage{graphicx}
\usepackage{amsfonts}
\usepackage{color}
\usepackage{bm}
\usepackage{xspace,soul}
\usepackage[breaklinks=true,colorlinks=true,urlcolor=black,linkcolor=blue,citecolor=red]{hyperref}
\usepackage{siunitx}
\usepackage{dsfont}

\usepackage{braket}
\usepackage{booktabs,multirow}
\graphicspath{{figures/}}

\usepackage{comment}
\usepackage{float}
\usepackage[vmargin=1in, hmargin=0.9in]{geometry}
\usepackage{times}
\usepackage[normalem]{ulem}

\begin{document}

\title{Deterministic correction of qubit loss}

\author{Roman Stricker}
\affiliation{Institut f\"ur Experimentalphysik, Universit\"at Innsbruck, Technikerstr.  25, A-6020 Innsbruck, Austria}
\author{Davide Vodola}
\affiliation{Department of Physics, College of Science, Swansea University, Singleton Park, SA2 8PP Swansea, United Kingdom}
\affiliation{Dipartimento di Fisica e Astronomia dell'Universit\`a di Bologna, and INFN, Sezione di Bologna, I-40127 Bologna, Italy}
\author{Alexander Erhard}
\affiliation{Institut f\"ur Experimentalphysik, Universit\"at Innsbruck, Technikerstr.  25, A-6020 Innsbruck, Austria}
\author{Lukas Postler}
\affiliation{Institut f\"ur Experimentalphysik, Universit\"at Innsbruck, Technikerstr.  25, A-6020 Innsbruck, Austria}
\author{Michael Meth}
\affiliation{Institut f\"ur Experimentalphysik, Universit\"at Innsbruck, Technikerstr.  25, A-6020 Innsbruck, Austria}
\author{Martin Ringbauer}
\affiliation{Institut f\"ur Experimentalphysik, Universit\"at Innsbruck, Technikerstr.  25, A-6020 Innsbruck, Austria}
\author{Philipp Schindler}
\affiliation{Institut f\"ur Experimentalphysik, Universit\"at Innsbruck, Technikerstr.  25, A-6020 Innsbruck, Austria}
\author{Thomas Monz}
\affiliation{Institut f\"ur Experimentalphysik, Universit\"at Innsbruck, Technikerstr.  25, A-6020 Innsbruck, Austria}
\affiliation{Alpine Quantum Technologies GmbH, 6020 Innsbruck, Austria}
\author{Markus M\"uller}
\affiliation{Department of Physics, College of Science, Swansea University, Singleton Park, SA2 8PP Swansea, United Kingdom}
\affiliation{Institute for Quantum Information, RWTH Aachen University, D-52056 Aachen, Germany}
\affiliation{Peter Gr\"unberg Institute, Theoretical Nanoelectronics, Forschungszentrum J\"ulich, D-52425 J\"ulich, Germany}
\author{Rainer Blatt}
\affiliation{Institut f\"ur Experimentalphysik, Universit\"at Innsbruck, Technikerstr.  25, A-6020 Innsbruck, Austria}
\affiliation{Institut  f\"ur  Quantenoptik  und  Quanteninformation, \"Osterreichische  Akademie  der  Wissenschaften, Otto-Hittmair-Platz  1, A-6020 Innsbruck, Austria}

\begin{abstract}
The loss of qubits -- the elementary carriers of quantum information -- poses one of the fundamental obstacles towards large-scale and fault-tolerant quantum information processors. In this work, we experimentally demonstrate a complete toolbox and the implementation of a full cycle of qubit loss detection and correction on a minimal instance of a topological surface code. This includes a quantum non-demolition measurement of a qubit loss event that conditionally triggers a restoration procedure, mapping the logical qubit onto a new encoding on the remaining qubits. The demonstrated methods, implemented here in a trapped-ion quantum processor, are applicable to other quantum computing architectures and codes, including leading 2D and 3D topological quantum error correcting codes. These tools complement previously demonstrated techniques to correct computational errors, and in combination constitute essential building blocks for complete and scalable quantum error correction.
\end{abstract}

\maketitle

Quantum error correction (QEC) \cite{PhysRevA.57.127} provides powerful techniques to detect and correct errors affecting quantum processors. Whereas most experimental efforts have thus far focused on correcting computational errors such as bit and phase flips \cite{Nigg302,Schindler1059, PhysRevLett.119.180501, Linkee1701074, Corcoles2015, Knill2001, Yao2012}, the loss of qubits from quantum registers \cite{PhysRevA.56.33} represents a fundamental, though often overlooked or neglected source of errors.

Qubit loss comes in a variety of physical incarnations such as the loss of particles encoding the qubits in atomic and photonic implementations \cite{PhysRevA.97.052301, lu-pnas-105-11050, Bell2014, 2058-9565-3-1-015005}, but also as leakage out of the two-dimensional computational-qubit subspace in multi-level solid-state \cite{PhysRevA.88.042308} and atomic, molecular, and optical systems \cite{PhysRevA.97.052301}. Whereas progress has been made in characterizing and suppressing the rate of loss and leakage processes \cite{PhysRevA.89.062321,PhysRevLett.114.100503, Kwon2017, PhysRevA.100.032325}, in many platforms these processes still occur at rates of the same order of magnitude as other errors, such as amplitude damping in trapped-ion qubits encoded in meta-stable states of optical transitions \cite{PhysRevA.97.052301}. It is known that unnoticed and uncorrected qubit loss and leakage will severely affect the performance of quantum processors \cite{PhysRevA.88.042308, PhysRevA.88.062329}, and thus dedicated protocols to fight this error source have been devised. These protocols include 4-qubit quantum erasure codes \cite{PhysRevA.56.33}, which have been implemented using photons and post-selective quantum state analysis \cite{lu-pnas-105-11050, Bell2014}, as well as protocols proposed to cope with qubit loss in the surface code \cite{Kitaev2003, Stace2009, varbanov2020} and 2D color codes \cite{Bombin2006, PhysRevLett.121.060501}. To date, an experimental implementation of deterministic detection and correction of qubit loss, however, remains an outstanding challenge.

%\section{Losses in the Kitaev surface code}
\label{ch:KitaevSurfaceCode}
\begin{figure*}[t]
\centering
\includegraphics[scale=0.9]{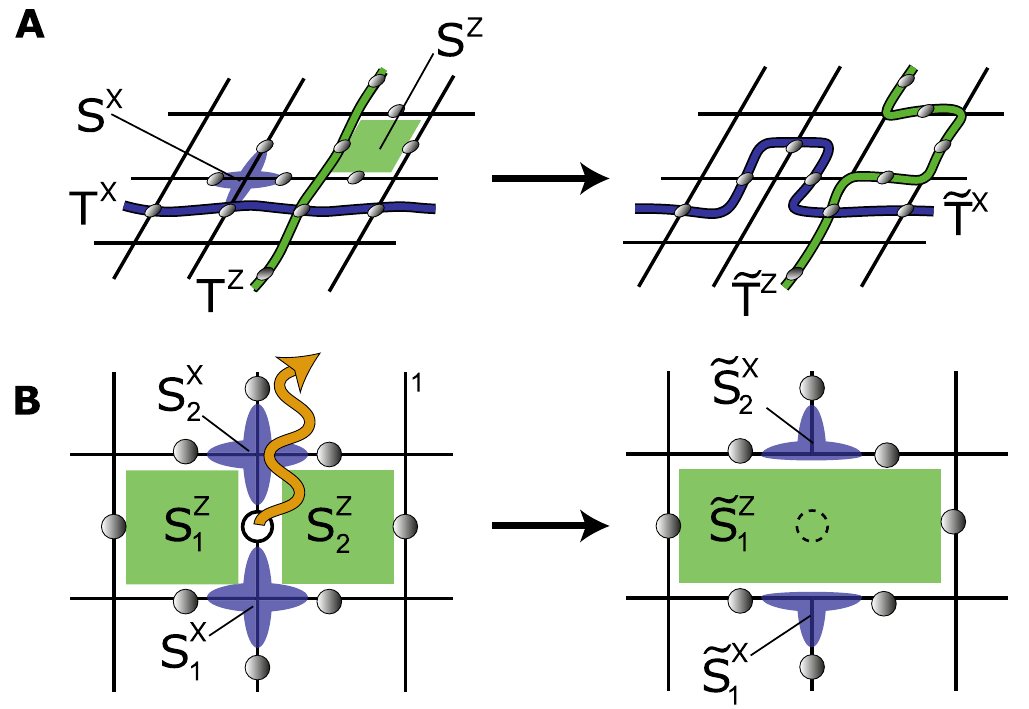}
\caption{\textbf{The surface code and correction of qubit loss}. (\textbf{A}) Logical qubits are encoded collectively in many physical qubits (grey circles) that are located on the edges of a 2D square lattice. The code space is defined via four-qubit $S^Z$ and $S^X$ stabilizers acting on groups of qubits residing around plaquettes (green square) and vertices (blue cross) of the lattice. Logical $T^Z$ and $T^X$ operators are defined along strings of qubits that span the entire lattice along two non-trivial paths, as depicted as the vertical green (horizontal blue) string for $T^Z$ ($T^X$). (Right panel) Logical string operators do not have unique support but can be deformed by multiplication with stabilizers, as illustrated for $T^Z$ ($T^X$) that is deformed into $\widetilde{T}^Z$ ($\widetilde{T}^X$) by the green plaquette (blue vertex) stabilizer. (\textbf{B}) (Left panel) Excerpt of the qubit lattice suffering the loss (orange arrow) of a physical qubit (white circle). The loss affects two plaquette operators $S^Z_1$ and $S^Z_2$ and two vertex operators $S^X_1$ and $S^X_2$. (Right panel) The correction algorithm consists of introducing a new merged $Z$ stabilizer generator as $\widetilde{S}^Z_1 =S^Z_1 S^Z_2$, which does not involve the lost qubit, and two new $X$ stabilizers $\widetilde{S}^X_1$, $\widetilde{S}^X_2$, both having reduced support on three qubits unaffected by the loss.}
\label{fig_kitaev_loss_lattice}
\end{figure*}

A general, architecture independent protocol to protect quantum information against loss errors consists in (i) the initial encoding of logical states into a multi-qubit register, (ii) a quantum non demolition (QND) measurement scheme that determines the position of potentially lost qubits, (iii) a reconstruction algorithm that, if not too many loss events have occurred, reconstructs the damaged code, and (iv) a final set of measurements that fixes the new code by initializing the new stabilizers.

In this work, we encode a single logical qubit in an excerpt of the surface code \cite{Kitaev2003,dennis-j-math-phys-43-4452}, which is a topological QEC code where physical qubits reside on the edges of a 2D square lattice, see Fig.~\ref{fig_kitaev_loss_lattice}A. The surface code is a Calderbank-Shor-Steane (CSS) code \cite{PhysRevA.54.1098, PhysRevLett.77.793}, for which stabilizer operators are associated to each vertex $V$ (blue cross in Fig.~\ref{fig_kitaev_loss_lattice}A) via $S^X_V = \prod_{j\in V} X_j$ and to each plaquette $P$ (green square in Fig.~\ref{fig_kitaev_loss_lattice}A) via $S^Z_P = \prod_{j\in P} Z_j$ where $X_j,Y_j,Z_j$ are Pauli matrices acting on the physical qubit $j$. All stabilizers mutually commute and their common $+1$ eigenspace fixes the code space that hosts the logical quantum states $\Ket{\psi_L}$, i.e.\ $S^Z_P |\psi_L\rangle = S^X_V |\psi_L\rangle = |\psi_L\rangle$ for all plaquettes and vertices. Operators that define and induce flips of the logical basis states $\ket{0_L}$ and $\ket{1_L}$ are the logical generators $T^Z$ and $T^X$, respectively. They commute with all stabilizers and can be chosen as products of $X$ and $Z$ operators along strings that span the entire lattice, see Fig.~\ref{fig_kitaev_loss_lattice}A. 

To recover a logical qubit affected by qubit loss, one needs to switch to an equivalent set of stabilizers $\{\widetilde{S}^X_V, \widetilde{S}^Z_P\}$ and logical operators $\{\widetilde{T}^X, \widetilde{T}^Z\}$ defined only on qubits that are not affected by losses. For this redefinition we follow the scheme introduced in Ref.~\cite{Stace2009} and shown in Fig.~\ref{fig_kitaev_loss_lattice}B. Notably, the logical operators do not have unique support as equivalent operators $\widetilde{T}^X$ and $\widetilde{T}^Z$ can be obtained by multiplying ${T}^X$ and ${T}^Z$ by any subset of stabilizers. For the surface code this results in the deformation of the string of physical qubits which supports the logical operator, see Fig.~\ref{fig_kitaev_loss_lattice}A. For too many losses, however, finding such an equivalent logical operator might not be possible. Since each loss event results in the deletion of one edge (bond) of the 2D square lattice, the question if such a path supporting a logical operator exists maps to the classical problem of bond percolation, which results for the surface code in a threshold of tolerable qubit loss rate in the absence of other errors as high as $50\%$ \cite{Stace2009}. 

\begin{figure*}[t!]
\includegraphics[width=1\textwidth]{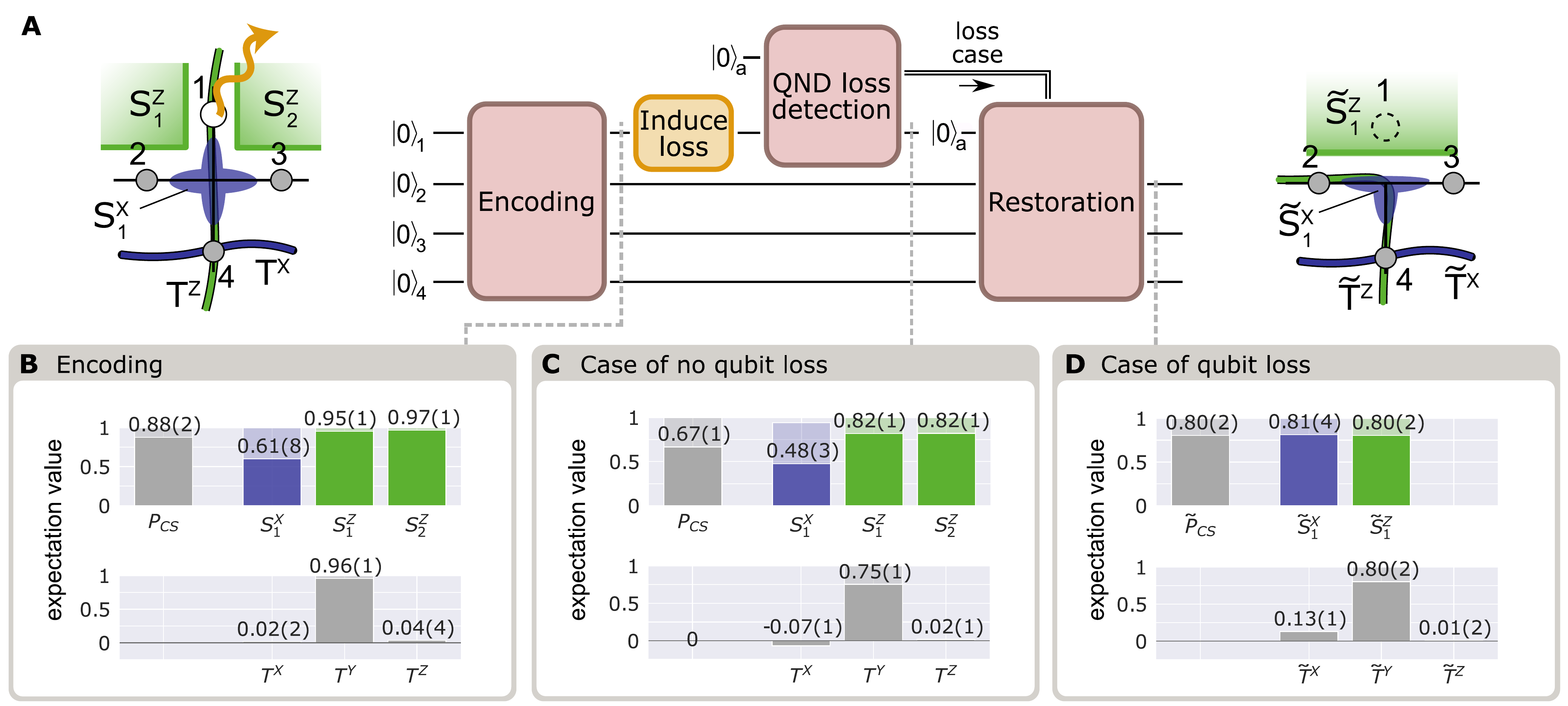}
\caption{\textbf{Experimental realization of the 1+4-qubit algorithm aiming at loss detection and correction.} (\textbf{A}) Minimal four-qubit system for the experimental realization of the full loss correction protocol. The code is defined by three stabilizers, $S^Z_1 = Z_1 Z_2$, $S^Z_2 = Z_1 Z_3$ (green squares) and $S^X_1 = X_1 X_2 X_3 X_4$ (blue cross) and stores a single logical qubit with logical operators $T^Z = Z_1 Z_4$,  $T^X = X_4$ and $T^Y = i T^X T^Z$. In the event of a loss (orange arrow) of qubit 1 (white circle), the merged $Z$ stabilizer $\widetilde{S}^Z_1 = S^Z_1 S^Z_2 = Z_2Z_3$ and a new $X$ stabilizer $\widetilde{S}^X_1 = X_2 X_3 X_4$ with reduced support on the remaining three qubits are introduced. The logical operators equivalent to the previous ones are $\widetilde{T}^Z = S^Z_1 T^Z = Z_2 Z_4$,  $\widetilde{T}^X = X_4$ and  $\widetilde{T}^Y =i \widetilde{T}^X \widetilde{T}^Z$.
(\textbf{B}) Expectation values for logical operators ($T$), stabilizers ($S$), and code space populations ($P_\text{CS}$) for the logical superposition state $\ket{+i_L}=(\ket{0_L}+i\ket{1_L})/\sqrt{2}$. All values are estimated from four-qubit quantum state tomography, with ideal values shaded in the background. (\textbf{C}) In the absence of loss, the logical encoding remains largely intact. (\textbf{D}) In case of loss, we reconstruct the code on the three remaining qubits after measuring the shrunk stabilizer $\widetilde{S}_1^{X} =X_2 X_3 X_4$ and selecting the appropriate Pauli basis, i.e.\ performing a Pauli frame update in the case of a $-1$ outcome in the $\widetilde{S}_1^{X}$ measurement.}
\label{fig_KitaevResults}
\end{figure*}

The minimal instance of the surface code that allows us to experimentally explore the reconstruction protocol consists of four physical qubits and is described in Fig.~\ref{fig_KitaevResults}A.
For the physical realization of this code we consider a string of $^{40}$Ca$^+$ ions confined in a linear Paul trap \cite{toolbox}. Each ion represents a physical qubit encoded in the electronic levels $S_{1/2}(m=-1/2) = \Ket{0}$ and $D_{5/2}(m=-1/2) = \Ket{1}$. Our setup is capable of realizing a universal set of quantum gate operations consisting in (a) single-qubit rotations by an angle $\theta$ around the z-axis of the form $\mathrm{R}^Z_j(\theta)=\exp(-i{\theta}{Z_j}/{2})$ on the $j${th} ion, (b) collective qubit rotations around the x- and the y-axes of the form $\mathrm{R}^\sigma(\theta) = \exp(-i{\theta}\sum_j {\sigma_j}/{2})$ with $\sigma = X$ or $Y$ via a laser beam addressing the entire register, and (c) multi-qubit M\o{}lmer-S\o{}rensen entangling gate operations $\mathrm{MS}^X(\theta) = \exp(-i{\theta}\sum_{j<\ell} X_j X_\ell /2)$ \cite{PhysRevLett.82.1835}. This gate set is complemented by single qubit hiding and unhiding operations in order to apply collective multi-qubit operations to only a subset of qubits \cite{toolbox}. Similarly this technique is used to read out individual qubits within the register without influencing the other qubits, see Appendix for details.

\begin{figure*}[h!t]
    \centering
    \includegraphics[width=0.85\textwidth]{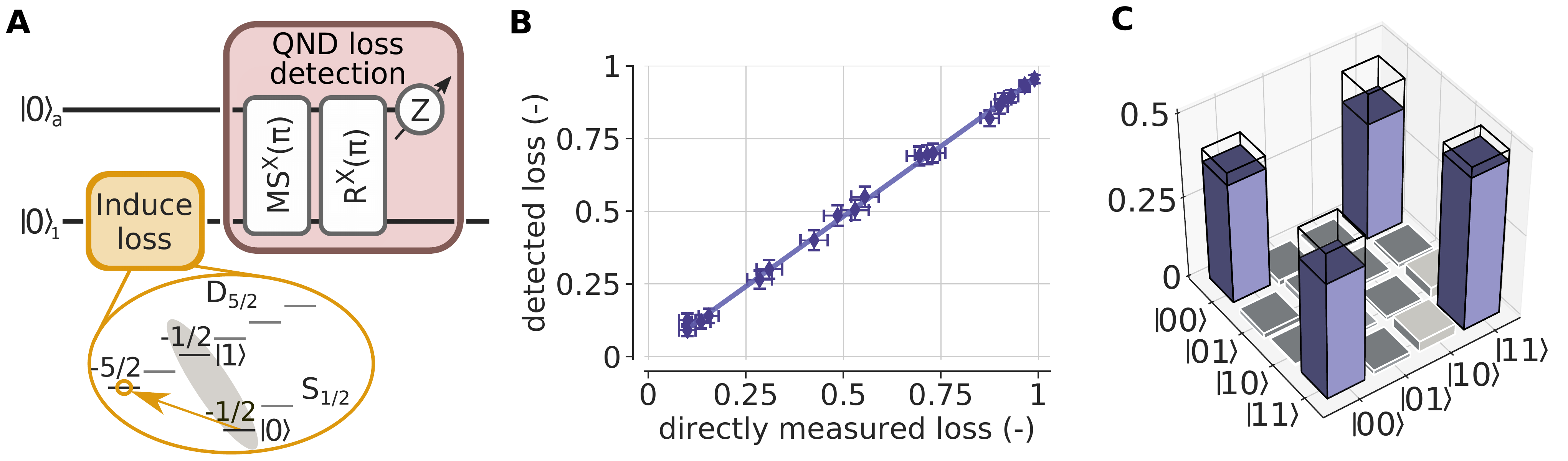}
\caption{\textbf{Investigating the performance of the QND loss detection unit.} (\textbf{A}) Circuit representation of the detection unit, mapping potential loss from qubit 1 onto the ancilla qubit. The following experimental results were extracted from experiments performed on the full 5-qubit register, according to Fig.~\ref{fig_KitaevResults}. (\textbf{B}) Population in the $D_{5/2}$-state of qubit 1 (directly measured loss) and ancilla qubit (detected loss) measured after loss detection. Controlled loss up to \SI{100}{\%} from state $\ket{0}$ was introduced. The estimated detection efficiency is \SI{96.5(4)}{\%}. This demonstrates that the occurrence of a loss event can be reliably mapped onto the ancilla qubit and read out in a QND fashion. (\textbf{C}) Reconstructed Choi matrix for a loss rate of $\sim\SI{20}{\%}$ ($\phi=\SI{0.3}{\pi}$) from the $|0\rangle$ state with a process-fidelity of \SI{90(2)}{\%}, compared to ideal values denoted by black frames. We find that the detection unit, as expected, performs a non-unitary evolution that deviates from the identity operator due to measurement back-action, see Appendix.}
\label{fig_DetectionResults}
\end{figure*}

In order to benchmark the performance of the protocol we will introduce qubit loss in a controlled way, see Fig.~\ref{fig_DetectionResults}A. The qubit potentially suffering a loss is partially pumped out of its computational subspace $\lbrace S_{1/2}(m=-1/2) = \Ket{0}$, $D_{5/2}(m=-1/2)=\Ket{1}\rbrace$ by coherently driving the carrier transition $S_{1/2}(m=-1/2)=\Ket{0}\leftrightarrow D_{5/2}(m=-5/2)=\Ket{2}$. In the following this is referred to as the loss operation $\mathrm{R_\text{loss}}(\phi)$ where the angle $\phi$ controls the probability of loss from the state $\Ket{0}$ via $p_\text{loss}\propto \sin^2(\phi/2)$, see Appendix for details.

To detect a loss event we implement a QND measurement as shown in Fig.~\ref{fig_DetectionResults}A, which signals the loss of a code qubit by a bit-flip on an ancillary qubit prepared in the state $\ket{0}$, followed by an addressed readout of the ancillary qubit. The key ingredient of this QND measurement is a two-qubit entangling gate operation $\mathrm{MS}^X(\pi)$ that performs a collective bit-flip operation on the code and ancilla qubits if the code qubit is present. If the code qubit has been lost, on the other hand, this operation acts only on the ancilla, on which it performs an identity operation, see Appendix. A subsequent collective bit-flip $\mathrm{R}^X(\pi)=X$ will flip the ancilla qubit to $\ket{1}$ before its addressed readout. In the case that no loss occurred, the collective bit-flip induced by $\mathrm{MS}^X(\pi)$ will be undone by the $\mathrm{R}^X(\pi)=X$ operation and the ancilla qubit will end in the state $\ket{0}$~\cite{toolbox}. The code qubit, on the other hand, will in this case undergo a non-unitary evolution given by (up to normalization) $\rho \mapsto E \rho E^\dagger$ with $E = \ket{1}\!\bra{1} + \cos (\phi/2) \ket{0}\!\bra{0}$, which for small loss rates ($\phi\sim 0$) converges to the identity operation. This is a consequence of the information gain via the ancilla measurement that no loss has incurred in this instance, see Appendix. 

We test the loss detection sub-circuit on the full 5-qubit register by driving the loss transition $\mathrm{R_\text{loss}}(\phi)$ on qubit 1 and measuring the population in the $D_{5/2}$-state on both code and ancilla qubit. This measurement does not distinguish between the different Zeeman sublevels of the $D_{5/2}$-state manifold. Figure~\ref{fig_DetectionResults}B shows that loss detected by the ancilla qubit matches the loss induced on qubit 1 within statistical uncertainty, indicating that a loss event is reliably detected. The quantified detection efficiency is \SI{96.5(4)}{\%}, with a false positive rate of 3$(\substack{+1 \\ -1})$~\% and a false negative rate of 1$(\substack{+1 \\ -1})$~\%. In order to quantify the performance of the QND detection scheme in the absence of loss, we reconstruct the Choi matrix \cite{choi1975} of the corresponding non-unitary map using generalized quantum process tomography. The reconstructed Choi matrix shown in Fig.~\ref{fig_DetectionResults}C confirms this dynamical behavior expected in the no loss case with a process fidelity of $\SI{90(2)}{\%}$ at a loss rate of $\sim\SI{20}{\%}$ ($\phi = \SI{0.3}{\pi}$). This demonstrates that information about loss on the code qubit can be reliably mapped onto the ancilla qubit. For general loss detection purposes, one could use the detection unit to probe all code qubits within the register sequentially.

To investigate the robustness of our minimal instance logical qubit against loss, we combine the loss detection unit and conditional correction step in a 1+4-qubit algorithm sketched in Fig.~\ref{fig_KitaevResults}A. The experimental sequence for encoding an arbitrary input state of the form ${\Ket{\psi_L} = \cos(\alpha/2)\Ket{0_L} + i\sin(\alpha/2)\Ket{1_L}}$ in our ion-trap quantum computer is given in the Appendix. The logical basis states $\Ket{0_L}$ and $\Ket{1_L}$ encoded by the initial stabilizers read ${\Ket{0_L} = (\Ket{0000}+\Ket{1111}})/\sqrt{2}$ and ${\Ket{1_L} = (\Ket{0001}+\Ket{1110}})/\sqrt{2}$. These GHZ-states are produced with a single fully-entangling MS-gate $\mathrm{MS}^X(\pi/2)$ acting on all four code qubits, supported by additional local operations. Loss is observed using the QND-detection unit utilizing an ancilla qubit for loss readout. In this smallest excerpt of the surface code we consider potential qubit loss to happen on qubit 1 only, and hence we only probe qubit 1 using the QND detection unit as indicated in Fig.~\ref{fig_KitaevResults}A. Conditional on the detection of a loss event, our control scheme triggers a real-time deterministic code restoration via feed-forward. If no loss is detected, the logical states can be verified by measuring the generators of the stabilizer group $\lbrace S^Z_1 = Z_1 Z_2, S^Z_2 = Z_1 Z_3, S^X_1 = X_1 X_2 X_3 X_4\rbrace$ as well as the logical operators $\lbrace T^Z = Z_1 Z_4, T^X = X_4, T^Y = i T^X T^Z\rbrace$ of the original encoding. If loss occurs, the encoded logical information can be restored by switching to an encoding defined on a smaller subset of three qubits. This is realized by a projective measurement of the shrunk stabilizer $\widetilde{S}_1^{X} =X_2 X_3 X_4$, which after the loss is in an undetermined state. This initializes the three-qubit stabilizer in a $+1$ (or $-1$) eigenstate, where the $-1$ case requires a redefinition of the Pauli basis (Pauli frame update) \cite{Knill2005, Aliferis:2006:QAT:2011665.2011666}, see Appendix for details. For this stabilizer readout, a freshly initialized ancilla qubit is needed. In our implementation we recycle the ancilla qubit, previously used for the QND loss detection, since it remains unaffected by the measurement in the loss case. Following this procedure, the initial logical encoding is reconstructed in the smaller subset of three qubits, see Fig.~\ref{fig_KitaevResults}A.

We now present the results obtained from the full implementation of the 1+4-qubit algorithm, as shown in Fig.~\ref{fig_KitaevResults}. Data was taken for three different input states, namely the logical basis states $\ket{0_L}$ and $\ket{1_L}$, presented in the Appendix, as well as their respective superposition $\ket{+i_L}=(\ket{0_L}+i\Ket{1_L})/\sqrt{2}$ presented here. To verify the initialization of $\ket{+i_L}$ we reconstruct the experimental density matrix via four-qubit quantum state tomography on the code qubits, yielding a fidelity of $\SI{85(1)}{\%}$ with the ideal state. 
From the reconstructed density matrix we further extract the components of the ``logical'' Bloch vector, represented by expectation values of the associated logical operators, the code space population $P_\text{CS}$, explained in the Appendix, and the expectation values of the stabilizer generators summarized in Fig.~\ref{fig_KitaevResults}B.

After the encoding, partial loss on qubit 1 is induced by coherently exciting the loss transition $\mathrm{R_\text{loss}}(\phi)$ for different values of $\phi$. Here, we present the case of \SI{25}{\%} loss, i.e.\ $\phi = \SI{0.5}{\pi}$, and other values are found in the Appendix. Loss is detected by a QND measurement mapping the information of loss onto the ancilla qubit, followed by a projective measurement of the ancilla qubit. The measurement result triggers a real-time deterministic code restoration via feed-forward. If no loss is detected quantum state tomography on all four code qubits is performed to verify the initial encoding $\ket{+i_L}$ to be still intact, with a fidelity of $\SI{67(1)}{\%}$ with the expected state, see Fig.~\ref{fig_KitaevResults}C. In case of detecting loss the code is switched to the remaining three qubits by a projective measurement of the shrunk stabilizer $\widetilde{S}_1^{X}$ as illustrated in Fig.~\ref{fig_KitaevResults}A and a Pauli frame update in case of a $-1$ outcome. Quantum state tomography yields a fidelity of the resulting three-qubit logical state $\ket{+i_L}$ of \SI{78(1)}{\%}, see Fig.~\ref{fig_KitaevResults}D.

The observed decrease in fidelity after loss detection is mainly due to cross-talk between neighboring ions resulting in unitary errors on the final state, and dephasing due to laser-frequency and magnetic-field fluctuations. Additionally in the no-loss case, the ancilla qubit has scattered photons during the in-sequence loss detection. This heats up the ion-string, decreasing the quality of the subsequent tomography operations.

Our work demonstrates the first deterministic detection and correction of qubit loss. Our building blocks are readily applicable to leading QEC codes such as the surface and color code and fully compatible with the framework of topological QEC. Whereas demonstrated here on an ion quantum processor, essentially all experimental quantum computing platforms are affected by qubit loss or leakage and could thus benefit from our methods. A fault-tolerant implementation of the presented routines in combination with correction of computational errors represents the next step towards large-scale quantum computers.

\bibliographystyle{apsrev}
\bibliography{biblio}

\begin{thebibliography}{33}
\expandafter\ifx\csname natexlab\endcsname\relax\def\natexlab#1{#1}\fi
\expandafter\ifx\csname bibnamefont\endcsname\relax
  \def\bibnamefont#1{#1}\fi
\expandafter\ifx\csname bibfnamefont\endcsname\relax
  \def\bibfnamefont#1{#1}\fi
\expandafter\ifx\csname citenamefont\endcsname\relax
  \def\citenamefont#1{#1}\fi
\expandafter\ifx\csname url\endcsname\relax
  \def\url#1{\texttt{#1}}\fi
\expandafter\ifx\csname urlprefix\endcsname\relax\def\urlprefix{URL }\fi
\providecommand{\bibinfo}[2]{#2}
\providecommand{\eprint}[2][]{\url{#2}}

\bibitem[{\citenamefont{Gottesman}(1998)}]{PhysRevA.57.127}
\bibinfo{author}{\bibfnamefont{D.}~\bibnamefont{Gottesman}},
  \bibinfo{journal}{Phys. Rev. A} \textbf{\bibinfo{volume}{57}},
  \bibinfo{pages}{127} (\bibinfo{year}{1998}).

\bibitem[{\citenamefont{Nigg et~al.}(2014)\citenamefont{Nigg, M{\"u}ller,
  Martinez, Schindler, Hennrich, Monz, Martin-Delgado, and Blatt}}]{Nigg302}
\bibinfo{author}{\bibfnamefont{D.}~\bibnamefont{Nigg}},
  \bibinfo{author}{\bibfnamefont{M.}~\bibnamefont{M{\"u}ller}},
  \bibinfo{author}{\bibfnamefont{E.~A.} \bibnamefont{Martinez}},
  \bibinfo{author}{\bibfnamefont{P.}~\bibnamefont{Schindler}},
  \bibinfo{author}{\bibfnamefont{M.}~\bibnamefont{Hennrich}},
  \bibinfo{author}{\bibfnamefont{T.}~\bibnamefont{Monz}},
  \bibinfo{author}{\bibfnamefont{M.~A.} \bibnamefont{Martin-Delgado}},
  \bibnamefont{and} \bibinfo{author}{\bibfnamefont{R.}~\bibnamefont{Blatt}},
  \bibinfo{journal}{Science} \textbf{\bibinfo{volume}{345}},
  \bibinfo{pages}{302} (\bibinfo{year}{2014}).

\bibitem[{\citenamefont{Schindler et~al.}(2011)\citenamefont{Schindler,
  Barreiro, Monz, Nebendahl, Nigg, Chwalla, Hennrich, and
  Blatt}}]{Schindler1059}
\bibinfo{author}{\bibfnamefont{P.}~\bibnamefont{Schindler}},
  \bibinfo{author}{\bibfnamefont{J.~T.} \bibnamefont{Barreiro}},
  \bibinfo{author}{\bibfnamefont{T.}~\bibnamefont{Monz}},
  \bibinfo{author}{\bibfnamefont{V.}~\bibnamefont{Nebendahl}},
  \bibinfo{author}{\bibfnamefont{D.}~\bibnamefont{Nigg}},
  \bibinfo{author}{\bibfnamefont{M.}~\bibnamefont{Chwalla}},
  \bibinfo{author}{\bibfnamefont{M.}~\bibnamefont{Hennrich}}, \bibnamefont{and}
  \bibinfo{author}{\bibfnamefont{R.}~\bibnamefont{Blatt}},
  \bibinfo{journal}{Science} \textbf{\bibinfo{volume}{332}},
  \bibinfo{pages}{1059} (\bibinfo{year}{2011}).

\bibitem[{\citenamefont{Takita et~al.}(2017)\citenamefont{Takita, Cross,
  C\'orcoles, Chow, and Gambetta}}]{PhysRevLett.119.180501}
\bibinfo{author}{\bibfnamefont{M.}~\bibnamefont{Takita}},
  \bibinfo{author}{\bibfnamefont{A.~W.} \bibnamefont{Cross}},
  \bibinfo{author}{\bibfnamefont{A.~D.} \bibnamefont{C\'orcoles}},
  \bibinfo{author}{\bibfnamefont{J.~M.} \bibnamefont{Chow}}, \bibnamefont{and}
  \bibinfo{author}{\bibfnamefont{J.~M.} \bibnamefont{Gambetta}},
  \bibinfo{journal}{Phys. Rev. Lett.} \textbf{\bibinfo{volume}{119}},
  \bibinfo{pages}{180501} (\bibinfo{year}{2017}).

\bibitem[{\citenamefont{Linke et~al.}(2017)\citenamefont{Linke, Gutierrez,
  Landsman, Figgatt, Debnath, Brown, and Monroe}}]{Linkee1701074}
\bibinfo{author}{\bibfnamefont{N.~M.} \bibnamefont{Linke}},
  \bibinfo{author}{\bibfnamefont{M.}~\bibnamefont{Gutierrez}},
  \bibinfo{author}{\bibfnamefont{K.~A.} \bibnamefont{Landsman}},
  \bibinfo{author}{\bibfnamefont{C.}~\bibnamefont{Figgatt}},
  \bibinfo{author}{\bibfnamefont{S.}~\bibnamefont{Debnath}},
  \bibinfo{author}{\bibfnamefont{K.~R.} \bibnamefont{Brown}}, \bibnamefont{and}
  \bibinfo{author}{\bibfnamefont{C.}~\bibnamefont{Monroe}},
  \bibinfo{journal}{Sci. Adv.} \textbf{\bibinfo{volume}{3}},
  \bibinfo{pages}{e1701074} (\bibinfo{year}{2017}).

\bibitem[{\citenamefont{C{\'o}rcoles et~al.}(2015)\citenamefont{C{\'o}rcoles,
  Magesan, Srinivasan, Cross, Steffen, Gambetta, and Chow}}]{Corcoles2015}
\bibinfo{author}{\bibfnamefont{A.~D.} \bibnamefont{C{\'o}rcoles}},
  \bibinfo{author}{\bibfnamefont{E.}~\bibnamefont{Magesan}},
  \bibinfo{author}{\bibfnamefont{S.~J.} \bibnamefont{Srinivasan}},
  \bibinfo{author}{\bibfnamefont{A.~W.} \bibnamefont{Cross}},
  \bibinfo{author}{\bibfnamefont{M.}~\bibnamefont{Steffen}},
  \bibinfo{author}{\bibfnamefont{J.~M.} \bibnamefont{Gambetta}},
  \bibnamefont{and} \bibinfo{author}{\bibfnamefont{J.~M.} \bibnamefont{Chow}},
  \bibinfo{journal}{Nat. Commun.} \textbf{\bibinfo{volume}{6}},
  \bibinfo{pages}{6979} (\bibinfo{year}{2015}).

\bibitem[{\citenamefont{Knill et~al.}(2001)\citenamefont{Knill, Laflamme,
  Martinez, and Negrevergne}}]{Knill2001}
\bibinfo{author}{\bibfnamefont{E.}~\bibnamefont{Knill}},
  \bibinfo{author}{\bibfnamefont{R.}~\bibnamefont{Laflamme}},
  \bibinfo{author}{\bibfnamefont{R.}~\bibnamefont{Martinez}}, \bibnamefont{and}
  \bibinfo{author}{\bibfnamefont{C.}~\bibnamefont{Negrevergne}},
  \bibinfo{journal}{Phys. Rev. Lett.} \textbf{\bibinfo{volume}{86}},
  \bibinfo{pages}{5811} (\bibinfo{year}{2001}).

\bibitem[{\citenamefont{Yao et~al.}(2012)\citenamefont{Yao, Wang, Chen, Gao,
  Fowler, Raussendorf, Chen, Liu, Lu, Deng et~al.}}]{Yao2012}
\bibinfo{author}{\bibfnamefont{X.-C.} \bibnamefont{Yao}},
  \bibinfo{author}{\bibfnamefont{T.-X.} \bibnamefont{Wang}},
  \bibinfo{author}{\bibfnamefont{H.-Z.} \bibnamefont{Chen}},
  \bibinfo{author}{\bibfnamefont{W.-B.} \bibnamefont{Gao}},
  \bibinfo{author}{\bibfnamefont{A.~G.} \bibnamefont{Fowler}},
  \bibinfo{author}{\bibfnamefont{R.}~\bibnamefont{Raussendorf}},
  \bibinfo{author}{\bibfnamefont{Z.-B.} \bibnamefont{Chen}},
  \bibinfo{author}{\bibfnamefont{N.-L.} \bibnamefont{Liu}},
  \bibinfo{author}{\bibfnamefont{C.-Y.} \bibnamefont{Lu}},
  \bibinfo{author}{\bibfnamefont{Y.-J.} \bibnamefont{Deng}},
  \bibnamefont{et~al.}, \bibinfo{journal}{Nature}
  \textbf{\bibinfo{volume}{482}}, \bibinfo{pages}{489} (\bibinfo{year}{2012}).

\bibitem[{\citenamefont{Grassl et~al.}(1997)\citenamefont{Grassl, Beth, and
  Pellizzari}}]{PhysRevA.56.33}
\bibinfo{author}{\bibfnamefont{M.}~\bibnamefont{Grassl}},
  \bibinfo{author}{\bibfnamefont{T.}~\bibnamefont{Beth}}, \bibnamefont{and}
  \bibinfo{author}{\bibfnamefont{T.}~\bibnamefont{Pellizzari}},
  \bibinfo{journal}{Phys. Rev. A} \textbf{\bibinfo{volume}{56}},
  \bibinfo{pages}{33} (\bibinfo{year}{1997}).

\bibitem[{\citenamefont{Brown and Brown}(2018)}]{PhysRevA.97.052301}
\bibinfo{author}{\bibfnamefont{N.~C.} \bibnamefont{Brown}} \bibnamefont{and}
  \bibinfo{author}{\bibfnamefont{K.~R.} \bibnamefont{Brown}},
  \bibinfo{journal}{Phys. Rev. A} \textbf{\bibinfo{volume}{97}},
  \bibinfo{pages}{052301} (\bibinfo{year}{2018}).

\bibitem[{\citenamefont{Lu et~al.}(2008)\citenamefont{Lu, Gao, Zhang, Zhou,
  Yang, and Pan}}]{lu-pnas-105-11050}
\bibinfo{author}{\bibfnamefont{C.-Y.} \bibnamefont{Lu}},
  \bibinfo{author}{\bibfnamefont{W.-B.} \bibnamefont{Gao}},
  \bibinfo{author}{\bibfnamefont{J.}~\bibnamefont{Zhang}},
  \bibinfo{author}{\bibfnamefont{X.-Q.} \bibnamefont{Zhou}},
  \bibinfo{author}{\bibfnamefont{T.}~\bibnamefont{Yang}}, \bibnamefont{and}
  \bibinfo{author}{\bibfnamefont{J.-W.} \bibnamefont{Pan}},
  \bibinfo{journal}{Proc. Natl. Acad. Sci.} \textbf{\bibinfo{volume}{105}},
  \bibinfo{pages}{11050} (\bibinfo{year}{2008}).

\bibitem[{\citenamefont{Bell et~al.}(2014)\citenamefont{Bell,
  Herrera-Mart{\'\i}, Tame, Markham, Wadsworth, and Rarity}}]{Bell2014}
\bibinfo{author}{\bibfnamefont{B.~A.} \bibnamefont{Bell}},
  \bibinfo{author}{\bibfnamefont{D.~A.} \bibnamefont{Herrera-Mart{\'\i}}},
  \bibinfo{author}{\bibfnamefont{M.~S.} \bibnamefont{Tame}},
  \bibinfo{author}{\bibfnamefont{D.}~\bibnamefont{Markham}},
  \bibinfo{author}{\bibfnamefont{W.~J.} \bibnamefont{Wadsworth}},
  \bibnamefont{and} \bibinfo{author}{\bibfnamefont{J.~G.}
  \bibnamefont{Rarity}}, \bibinfo{journal}{Nature Communications}
  \textbf{\bibinfo{volume}{5}}, \bibinfo{pages}{3658} (\bibinfo{year}{2014}).

\bibitem[{\citenamefont{Morley-Short et~al.}(2018)\citenamefont{Morley-Short,
  Bartolucci, Gimeno-Segovia, Shadbolt, Cable, and
  Rudolph}}]{2058-9565-3-1-015005}
\bibinfo{author}{\bibfnamefont{S.}~\bibnamefont{Morley-Short}},
  \bibinfo{author}{\bibfnamefont{S.}~\bibnamefont{Bartolucci}},
  \bibinfo{author}{\bibfnamefont{M.}~\bibnamefont{Gimeno-Segovia}},
  \bibinfo{author}{\bibfnamefont{P.}~\bibnamefont{Shadbolt}},
  \bibinfo{author}{\bibfnamefont{H.}~\bibnamefont{Cable}}, \bibnamefont{and}
  \bibinfo{author}{\bibfnamefont{T.}~\bibnamefont{Rudolph}},
  \bibinfo{journal}{Quant. Sci. Tech.} \textbf{\bibinfo{volume}{3}},
  \bibinfo{pages}{015005} (\bibinfo{year}{2018}).

\bibitem[{\citenamefont{Fowler}(2013)}]{PhysRevA.88.042308}
\bibinfo{author}{\bibfnamefont{A.~G.} \bibnamefont{Fowler}},
  \bibinfo{journal}{Phys. Rev. A} \textbf{\bibinfo{volume}{88}},
  \bibinfo{pages}{042308} (\bibinfo{year}{2013}).

\bibitem[{\citenamefont{Epstein et~al.}(2014)\citenamefont{Epstein, Cross,
  Magesan, and Gambetta}}]{PhysRevA.89.062321}
\bibinfo{author}{\bibfnamefont{J.~M.} \bibnamefont{Epstein}},
  \bibinfo{author}{\bibfnamefont{A.~W.} \bibnamefont{Cross}},
  \bibinfo{author}{\bibfnamefont{E.}~\bibnamefont{Magesan}}, \bibnamefont{and}
  \bibinfo{author}{\bibfnamefont{J.~M.} \bibnamefont{Gambetta}},
  \bibinfo{journal}{Phys. Rev. A} \textbf{\bibinfo{volume}{89}},
  \bibinfo{pages}{062321} (\bibinfo{year}{2014}).

\bibitem[{\citenamefont{Xia et~al.}(2015)\citenamefont{Xia, Lichtman, Maller,
  Carr, Piotrowicz, Isenhower, and Saffman}}]{PhysRevLett.114.100503}
\bibinfo{author}{\bibfnamefont{T.}~\bibnamefont{Xia}},
  \bibinfo{author}{\bibfnamefont{M.}~\bibnamefont{Lichtman}},
  \bibinfo{author}{\bibfnamefont{K.}~\bibnamefont{Maller}},
  \bibinfo{author}{\bibfnamefont{A.~W.} \bibnamefont{Carr}},
  \bibinfo{author}{\bibfnamefont{M.~J.} \bibnamefont{Piotrowicz}},
  \bibinfo{author}{\bibfnamefont{L.}~\bibnamefont{Isenhower}},
  \bibnamefont{and} \bibinfo{author}{\bibfnamefont{M.}~\bibnamefont{Saffman}},
  \bibinfo{journal}{Phys. Rev. Lett.} \textbf{\bibinfo{volume}{114}},
  \bibinfo{pages}{100503} (\bibinfo{year}{2015}).

\bibitem[{\citenamefont{Kwon et~al.}(2017)\citenamefont{Kwon, Ebert, Walker,
  and Saffman}}]{Kwon2017}
\bibinfo{author}{\bibfnamefont{M.}~\bibnamefont{Kwon}},
  \bibinfo{author}{\bibfnamefont{M.~F.} \bibnamefont{Ebert}},
  \bibinfo{author}{\bibfnamefont{T.~G.} \bibnamefont{Walker}},
  \bibnamefont{and} \bibinfo{author}{\bibfnamefont{M.}~\bibnamefont{Saffman}},
  \bibinfo{journal}{Phys. Rev. Lett.} \textbf{\bibinfo{volume}{119}},
  \bibinfo{pages}{180504} (\bibinfo{year}{2017}).

\bibitem[{\citenamefont{Brown and Brown}(2019)}]{PhysRevA.100.032325}
\bibinfo{author}{\bibfnamefont{N.~C.} \bibnamefont{Brown}} \bibnamefont{and}
  \bibinfo{author}{\bibfnamefont{K.~R.} \bibnamefont{Brown}},
  \bibinfo{journal}{Phys. Rev. A} \textbf{\bibinfo{volume}{100}},
  \bibinfo{pages}{032325} (\bibinfo{year}{2019}).

\bibitem[{\citenamefont{Ghosh et~al.}(2013)\citenamefont{Ghosh, Fowler,
  Martinis, and Geller}}]{PhysRevA.88.062329}
\bibinfo{author}{\bibfnamefont{J.}~\bibnamefont{Ghosh}},
  \bibinfo{author}{\bibfnamefont{A.~G.} \bibnamefont{Fowler}},
  \bibinfo{author}{\bibfnamefont{J.~M.} \bibnamefont{Martinis}},
  \bibnamefont{and} \bibinfo{author}{\bibfnamefont{M.~R.}
  \bibnamefont{Geller}}, \bibinfo{journal}{Phys. Rev. A}
  \textbf{\bibinfo{volume}{88}}, \bibinfo{pages}{062329}
  (\bibinfo{year}{2013}).

\bibitem[{\citenamefont{Kitaev}(2003)}]{Kitaev2003}
\bibinfo{author}{\bibfnamefont{A.}~\bibnamefont{Kitaev}},
  \bibinfo{journal}{Ann. Phys.} \textbf{\bibinfo{volume}{303}},
  \bibinfo{pages}{2 } (\bibinfo{year}{2003}).

\bibitem[{\citenamefont{Stace et~al.}(2009)\citenamefont{Stace, Barrett, and
  Doherty}}]{Stace2009}
\bibinfo{author}{\bibfnamefont{T.~M.} \bibnamefont{Stace}},
  \bibinfo{author}{\bibfnamefont{S.~D.} \bibnamefont{Barrett}},
  \bibnamefont{and} \bibinfo{author}{\bibfnamefont{A.~C.}
  \bibnamefont{Doherty}}, \bibinfo{journal}{Phys. Rev. Lett.}
  \textbf{\bibinfo{volume}{102}}, \bibinfo{pages}{200501}
  (\bibinfo{year}{2009}).

\bibitem[{\citenamefont{Varbanov et~al.}((2020))\citenamefont{Varbanov,
  Battistel, Tarasinski, Ostroukh, O'Brien, DiCarlo, and
  Terhal}}]{varbanov2020}
\bibinfo{author}{\bibfnamefont{B.~M.} \bibnamefont{Varbanov}},
  \bibinfo{author}{\bibfnamefont{F.}~\bibnamefont{Battistel}},
  \bibinfo{author}{\bibfnamefont{B.~M.} \bibnamefont{Tarasinski}},
  \bibinfo{author}{\bibfnamefont{V.~P.} \bibnamefont{Ostroukh}},
  \bibinfo{author}{\bibfnamefont{T.~E.} \bibnamefont{O'Brien}},
  \bibinfo{author}{\bibfnamefont{L.}~\bibnamefont{DiCarlo}}, \bibnamefont{and}
  \bibinfo{author}{\bibfnamefont{B.~M.} \bibnamefont{Terhal}},
  \emph{\bibinfo{title}{Leakage detection for a transmon-based surface code}}
  (\bibinfo{year}{(2020)}), \eprint{arXiv:2002.07119}.

\bibitem[{\citenamefont{Bombin and Martin-Delgado}(2006)}]{Bombin2006}
\bibinfo{author}{\bibfnamefont{H.}~\bibnamefont{Bombin}} \bibnamefont{and}
  \bibinfo{author}{\bibfnamefont{M.~A.} \bibnamefont{Martin-Delgado}},
  \bibinfo{journal}{Phys. Rev. Lett.} \textbf{\bibinfo{volume}{97}},
  \bibinfo{pages}{180501} (\bibinfo{year}{2006}).

\bibitem[{\citenamefont{Vodola et~al.}(2018)\citenamefont{Vodola, Amaro,
  Martin-Delgado, and M\"uller}}]{PhysRevLett.121.060501}
\bibinfo{author}{\bibfnamefont{D.}~\bibnamefont{Vodola}},
  \bibinfo{author}{\bibfnamefont{D.}~\bibnamefont{Amaro}},
  \bibinfo{author}{\bibfnamefont{M.~A.} \bibnamefont{Martin-Delgado}},
  \bibnamefont{and} \bibinfo{author}{\bibfnamefont{M.}~\bibnamefont{M\"uller}},
  \bibinfo{journal}{Phys. Rev. Lett.} \textbf{\bibinfo{volume}{121}},
  \bibinfo{pages}{060501} (\bibinfo{year}{2018}).

\bibitem[{\citenamefont{Dennis et~al.}(2002)\citenamefont{Dennis, Kitaev,
  Landahl, and J}}]{dennis-j-math-phys-43-4452}
\bibinfo{author}{\bibfnamefont{E.}~\bibnamefont{Dennis}},
  \bibinfo{author}{\bibfnamefont{A.}~\bibnamefont{Kitaev}},
  \bibinfo{author}{\bibfnamefont{A.}~\bibnamefont{Landahl}}, \bibnamefont{and}
  \bibinfo{author}{\bibfnamefont{P.}~\bibnamefont{J}}, \bibinfo{journal}{J.
  Math. Phys.} \textbf{\bibinfo{volume}{43}}, \bibinfo{pages}{4452}
  (\bibinfo{year}{2002}).

\bibitem[{\citenamefont{Calderbank and Shor}(1996)}]{PhysRevA.54.1098}
\bibinfo{author}{\bibfnamefont{A.~R.} \bibnamefont{Calderbank}}
  \bibnamefont{and} \bibinfo{author}{\bibfnamefont{P.~W.} \bibnamefont{Shor}},
  \bibinfo{journal}{Phys. Rev. A} \textbf{\bibinfo{volume}{54}},
  \bibinfo{pages}{1098} (\bibinfo{year}{1996}).

\bibitem[{\citenamefont{Steane}(1996)}]{PhysRevLett.77.793}
\bibinfo{author}{\bibfnamefont{A.~M.} \bibnamefont{Steane}},
  \bibinfo{journal}{Phys. Rev. Lett.} \textbf{\bibinfo{volume}{77}},
  \bibinfo{pages}{793} (\bibinfo{year}{1996}).

\bibitem[{\citenamefont{Schindler et~al.}(2013)\citenamefont{Schindler, Nigg,
  Monz, Barreiro, Martinez, Wang, Quint, Brandl, Nebendahl, Roos
  et~al.}}]{toolbox}
\bibinfo{author}{\bibfnamefont{P.}~\bibnamefont{Schindler}},
  \bibinfo{author}{\bibfnamefont{D.}~\bibnamefont{Nigg}},
  \bibinfo{author}{\bibfnamefont{T.}~\bibnamefont{Monz}},
  \bibinfo{author}{\bibfnamefont{J.~T.} \bibnamefont{Barreiro}},
  \bibinfo{author}{\bibfnamefont{E.}~\bibnamefont{Martinez}},
  \bibinfo{author}{\bibfnamefont{S.~X.} \bibnamefont{Wang}},
  \bibinfo{author}{\bibfnamefont{S.}~\bibnamefont{Quint}},
  \bibinfo{author}{\bibfnamefont{M.~F.} \bibnamefont{Brandl}},
  \bibinfo{author}{\bibfnamefont{V.}~\bibnamefont{Nebendahl}},
  \bibinfo{author}{\bibfnamefont{C.~F.} \bibnamefont{Roos}},
  \bibnamefont{et~al.}, \bibinfo{journal}{New J. Phys.}
  \textbf{\bibinfo{volume}{15}}, \bibinfo{pages}{123012}
  (\bibinfo{year}{2013}).

\bibitem[{\citenamefont{M\o{}lmer and S\o{}rensen}(1999)}]{PhysRevLett.82.1835}
\bibinfo{author}{\bibfnamefont{K.}~\bibnamefont{M\o{}lmer}} \bibnamefont{and}
  \bibinfo{author}{\bibfnamefont{A.}~\bibnamefont{S\o{}rensen}},
  \bibinfo{journal}{Phys. Rev. Lett.} \textbf{\bibinfo{volume}{82}},
  \bibinfo{pages}{1835} (\bibinfo{year}{1999}).

\bibitem[{\citenamefont{Choi}(1975)}]{choi1975}
\bibinfo{author}{\bibfnamefont{M.-D.} \bibnamefont{Choi}},
  \bibinfo{journal}{Linear Algebra and its Applications}
  \textbf{\bibinfo{volume}{10}}, \bibinfo{pages}{285 } (\bibinfo{year}{1975}).

\bibitem[{\citenamefont{Knill}(2005)}]{Knill2005}
\bibinfo{author}{\bibfnamefont{E.}~\bibnamefont{Knill}},
  \bibinfo{journal}{Nature} \textbf{\bibinfo{volume}{434}}, \bibinfo{pages}{39}
  (\bibinfo{year}{2005}).

\bibitem[{\citenamefont{Aliferis et~al.}(2006)\citenamefont{Aliferis,
  Gottesman, and Preskill}}]{Aliferis:2006:QAT:2011665.2011666}
\bibinfo{author}{\bibfnamefont{P.}~\bibnamefont{Aliferis}},
  \bibinfo{author}{\bibfnamefont{D.}~\bibnamefont{Gottesman}},
  \bibnamefont{and} \bibinfo{author}{\bibfnamefont{J.}~\bibnamefont{Preskill}},
  \bibinfo{journal}{Quant. Info. Comput.} \textbf{\bibinfo{volume}{6}},
  \bibinfo{pages}{97} (\bibinfo{year}{2006}).

\bibitem[{\citenamefont{Nielsen and Chuang}(2011)}]{Nielsen:2011:QCQ:1972505}
\bibinfo{author}{\bibfnamefont{M.~A.} \bibnamefont{Nielsen}} \bibnamefont{and}
  \bibinfo{author}{\bibfnamefont{I.~L.} \bibnamefont{Chuang}},
  \emph{\bibinfo{title}{Quantum Computation and Quantum Information: 10th
  Anniversary Edition}} (\bibinfo{publisher}{Cambridge University Press},
  \bibinfo{address}{New York, USA}, \bibinfo{year}{2011}),
  \bibinfo{edition}{10th} ed.

\end{thebibliography}

\section*{Acknowledgements}
\textbf{Funding} We gratefully acknowledge funding by the U.S. Army Research Office (ARO) through grant no. W911NF-14-1-0103. We also acknowledge funding by the Austrian Science Fund (FWF), through the SFB BeyondC (FWF Project No. F71), by the Austrian Research Promotion Agency (FFG) contract 872766, by the EU H2020-FETFLAG-2018-03 under Grant Agreement no. 820495, and by the Office of the Director of National Intelligence (ODNI), Intelligence Advanced Research Projects Activity (IARPA), via the U.S. ARO Grant No. W911NF-16-1-0070. We acknowledge support from the Samsung Advanced Institute of Technology Global Research Outreach. This project has received funding from the European Union's Horizon 2020 research and innovation programme under the Marie Sk{\l}odowska-Curie grant agreement No. 801110 and the Austrian Federal Ministry of Education, Science and Research (BMBWF). It reflects only the author's view, the EU Agency is not responsible for any use that may be made of the information it contains.\\
\textbf{Author contributions} DV and MM derived the theory results. RS, AE, LP, MiM, MR, PS, and TM performed the experiments. RS analyzed the data. TM, MM, and RB supervised the project. All authors contributed to writing the manuscript. \\
\textbf{Competing interests} The authors declare no competing interests.

\setcounter{figure}{0}
\setcounter{equation}{0}
\setcounter{table}{0}
\makeatletter 
\renewcommand{\theequation}{S\@arabic\c@equation}
\renewcommand{\thefigure}{S\@arabic\c@figure}
\renewcommand{\thetable}{S\@arabic\c@table}
\makeatother

\onecolumngrid

\begin{center}
{\bf \large Appendix: \\
Deterministic correction of qubit loss}
\end{center}
\medskip

Here we provide further experimental and theoretical results and details on the detection and correction of qubit loss. We start in Sec.~\ref{supp:quantum-circuit} by presenting the quantum circuit specifically tailored for the toolbox given by our ion-trap quantum computer. We continue in Sec.~\ref{supp:hiding-unhiding} by explaining our approach to hide certain ions from the dynamics of collective M\o{}lmer-S\o{}renson entangling gates as well as collective readout operations by shelving their population in Zeeman sublevels outside the computational subspace. In Sec.~\ref{supp:QND-detection} we discuss the effective dynamics of the QND qubit loss detection scheme and deliver experimental data characterizing these dynamics. In Sec.~\ref{supp:additional-data} we provide complementary results on the full 1+4-qubit detection and correction algorithm for a larger number of logical input states and for in total three different qubit loss rates. In Sec.~\ref{supp:imperfections-QND} we present a model, which accounts for dominant experimental imperfections in the QND loss detection circuit and discuss how these limit the performance for current system parameters in the regime of low qubit loss rates. 

\section{Circuit representation of the 1+4 qubit loss detection and correction algorithm}
\label{supp:quantum-circuit}
The smallest instance for implementing a correction from qubit losses in the surface code is defined by four physical qubits forming a logical qubit in one plaquette as shown Fig~\ref{fig_KitaevResults}A. An additional ancilla qubit is required for QND loss detection. This leads to the 1+4-qubit loss detection and correction algorithm under study in the main text. In our detection and correction protocol, loss is considered to happen on qubit 1 only.

\begin{figure*}[ht]
    \centering
    \includegraphics[width=0.8\textwidth]{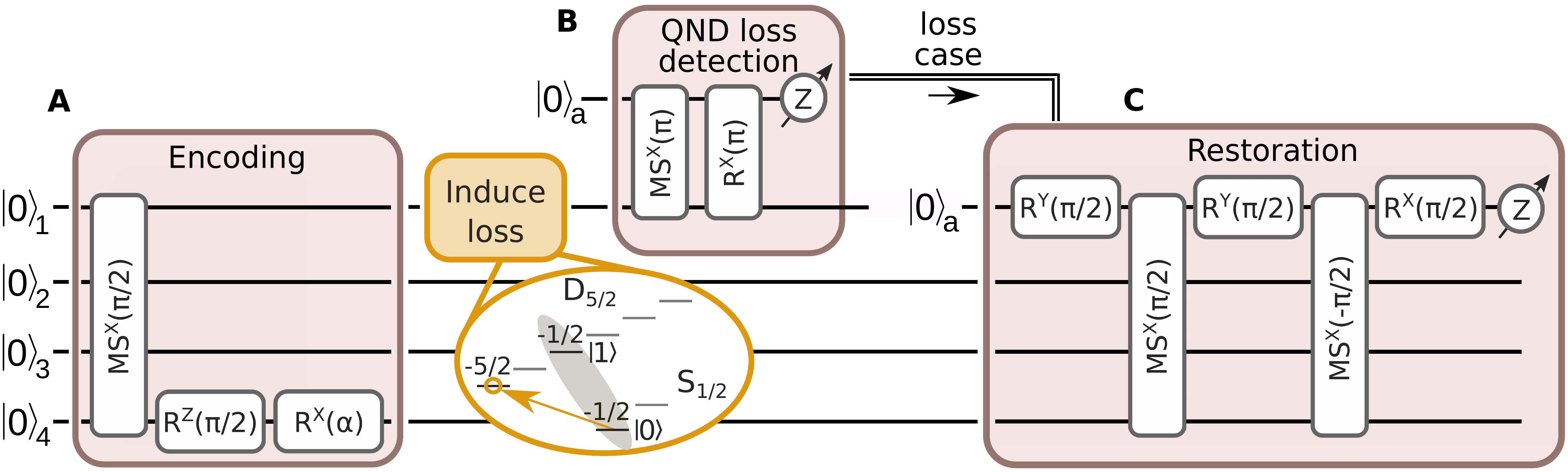}%
    \caption{\textbf{Gate sequence of the 1+4-qubit loss detection and correction algorithm.} (\textbf{A}) Encoding sequence implementing the smallest excerpt of Kitaev's surface code employing 4 physical qubits. Logical states of the form $\Ket{\psi_L}=\cos(\alpha/2)\Ket{0_L}+i\sin(\alpha/2)\Ket{1_L}$ with ${\Ket{0_L} = (\Ket{0000}+\Ket{1111}})/\sqrt{2}$ and ${\Ket{1_L} = (\Ket{0001}+\Ket{1110}})/\sqrt{2}$ are encoded and loss is induced in a controlled fashion on qubit 1. (\textbf{B}) QND detection unit identifying potential loss events on qubit 1. Conditional on the loss detection, our control scheme either keeps the original code or triggers a real-time deterministic code restoration via feed-forward. (\textbf{C}) The depicted gate sequence aims at measuring the shrunk stabilizer $\widetilde{S}_1^{X} =X_2 X_3 X_4$ to reconstruct the code in the smaller subset of the remaining three qubits. For this purpose we reuse the ancilla qubit from the detection-circuit, since it remains unaffected by the measurement in the loss case.}
    \label{fig_KitaevCircuit}
\end{figure*}

The related 5-qubit gate sequence, optimized for our ion-trap quantum computer, to encode an arbitrary logical input state of the form ${\Ket{\psi_L} = \cos(\alpha/2)\Ket{0_L} + i\sin(\alpha/2)\Ket{1_L}}$ is depicted in Fig~\ref{fig_KitaevCircuit}A. In our experimental toolbox the M\o{}lmer-S\o{}renson entangling gate operations $\mathrm{MS}^X(\theta) = \exp(-i{\theta}\sum_{j<\ell} X_j X_\ell /2)$ \cite{PhysRevLett.82.1835} acts as the entangling gate for multi-qubit operations. A fully-entangling gate $\mathrm{MS}^X(\pi/2)$, acting on all four code qubits, alongside local operations on qubit 4 lead to the GHZ-type logical basis states $\Ket{0_L} = (\Ket{0000}+\Ket{1111})/\sqrt{2}$ and $\Ket{1_L} = (\Ket{0001}+\Ket{1110})/\sqrt{2}$.

The subsequent QND loss detection unit in part B of Fig~\ref{fig_KitaevCircuit} combines a 2-qubit $\mathrm{MS}^X(\pi)$ followed by a collecive bit-flip $\mathrm{R}^X(\pi)=X$. In the absence of loss the $\mathrm{MS}^X(\pi)$ performs a bit-flip on both qubits present in the detection scheme. Whenever qubit 1 is outside the computational subspace, i.e. loss occurs, the MS-gate couples only to the ancilla qubit performing an identity operation, as can be seen from the argument of the exponential $\sum_{j<\ell} X_j X_\ell = X_j X_j = I$ in the above definition of the MS-gate. The subsequent $X$ operation flips the state of the ancilla qubit to $\Ket{1}$ followed by its addressed readout signaling the event of loss. If no loss was detected both gates add up to an overall identity operation leaving the logical encoding unaffected. In this way information about loss is mapped onto the ancilla qubit, which can be read out without influencing the logical encoding. Consequently, the described unit works in a quantum non demolition (QND) way. By probing all code qubits sequentially, one could extend this protocol to check the entire register for loss.

In the absence of loss, the logical encoding remains intact and can be verified by measuring the generators of the stabilizer group $\lbrace S^Z_1 = Z_1 Z_2, S^Z_2 = Z_1 Z_3, S^X_1 = X_1 X_2 X_3 X_4\rbrace$ as well as the logical operators $\lbrace T^Z = Z_1 Z_4, T^X = X_4, T^Y = i T^X T^Z\rbrace$ of the original encoding. If loss is detected on qubit 1, the encoded logical information can be restored by switching to an encoding defined on a smaller subset of three qubits, see Fig.~\ref{fig_KitaevResults}A. The merged $Z$ stabilizer $\widetilde{S}^Z_1 = S^Z_1 S^Z_2 = Z_2Z_3$ and a new $X$ stabilizer $\widetilde{S}^X_1 = X_2 X_3 X_4$ are introduced. This newly defined shrunk stabilizer $\widetilde{S}_1^{X} =X_2 X_3 X_4$ is, after the loss of qubit 1, in an undetermined state and needs to be measured to initialize the stabilizer in a +1 (or -1) eigenstate. Here, the -1 case requires a redefinition of the Pauli basis, as so called Pauli frame update \cite{Knill2005, Aliferis:2006:QAT:2011665.2011666}. The respective gate sequence mapping the syndrome onto the ancilla qubit, which is then read out, is shown in Fig~\ref{fig_KitaevCircuit}C. For this purpose we reuse the ancilla qubit from the QND detection unit, since it remains unaffected by the projective measurement in the loss case. Finally the logcial encoding is restored in a new encoding defined by the remaining three qubits. 

\section{Spectroscopic decoupling and recoupling of ions}
\label{supp:hiding-unhiding}
The circuit for the QND loss detection, depicted in Fig.~\ref{fig_KitaevCircuit}B, requires a 2-qubit entangling operation $\mathrm{MS}^X(\pi)$. This entangling gate operation is performed by a collective laser beam illuminating the entire ion string. However, these operations can be applied to a subset of qubits, by temporarily shelving the electronic populations of qubits not taking part in Zeeman sublevels outside the computational subspace. More precisely, population from the lower qubit state  $S_{1/2}(m=-1/2) = \Ket{0}$ is spectroscopically decoupled to $D_{5/2}(m=+1/2)$ and population from the upper qubit state $D_{5/2}(m=-1/2) = \Ket{1}$ is spectroscopically decoupled to $S_{1/2}(m=+1/2)$. In the main text we refer to this as hiding and unhiding operations. The same technique can be applied to read out individual qubits within the register without influencing the other qubits. Such addressed readout of (ancilla) qubits is essential to allow us to detect a qubit loss event and trigger a subsequent correction step via feed-forward.

\section{QND loss detection}
\label{supp:QND-detection}
This section begins with providing theoretical details of the protocol that introduces the loss and on the QND loss detection. We then also present additional experimental data characterizing the QND loss detection. 

The controlled loss operation on a code qubit ($q$) is realized by coherently transferring qubit population partially from the computational subspace spanned by $\{\ket{0} = S_{1/2}(m=-1/2)$ and $\ket{1} = D_{5/2}(m=-1/2)\}$ into the state $\ket{2} =  D_{5/2}(m=-5/2)$ via a coherent rotation $\mathrm{R_\text{loss}}(\phi)$ 
\begin{equation}
\mathrm{R_\text{loss}}(\phi) = \ket{1}\bra{1}_q + \cos \frac{\phi}{2} \left( \ket{0}\bra{0}_q + \ket{2}\bra{2}_q \right) + \sin \frac{\phi}{2} \left( \ket{0}\bra{2}_q - \ket{2}\bra{0}_q \right).
\end{equation}

The QND loss detection is realized by the circuit shown in Fig.~\ref{fig_KitaevCircuit}B. It consists of an MS-gate operation $\mathrm{MS}^X(\pi)$ between the code qubit ($q$) and the ancilla qubit ($a$) initially prepared in $\ket{0}$, followed by single-qubit bit flips $\mathrm{R^X}(\pi)$ applied to both the ancilla and the code qubit, and a projective measurement of the ancilla qubit in the computational basis. The two-qubit MS-gate applied to the data qubit ($q$) and the ancilla qubit ($a$) realizes the unitary
\begin{equation}
\mathrm{MS}^X(\phi) = \exp \left( - i\frac{\phi}{2} X_a X_q \right) = \left[ \cos \left( \frac{\phi}{2} \right) (1 - \ket{2}\bra{2}_q) - i \sin \left( \frac{\phi}{2} \right) X_a X_q\right] + \ket{2}\bra{2}_q,
\end{equation}
which is generated by $X_{i} = \ket{0}\bra{1}_{i} + \ket{0}\bra{0}_{i}$, for $i = q, a$, respectively, and reduces for $\phi=\pi$ to $\mathrm{MS}^X(\pi) = \ket{2}\bra{2}_q - i X_a X_q$.
Note that if both the code qubit $q$ and the ancilla qubit $a$ are initially in the computational subspace, this two-qubit operation realizes a collective bit flip (within the computational subspace). In contrast, if the code qubit is in $\ket{2}$, i.e.~outside the computational subspace, the state of the code and ancilla qubit remains unchanged under this operation~\cite{PhysRevLett.82.1835}.

The subsequent single-qubit rotations $\mathrm{R^X}(\pi)$ (bit flips) on both the data and the ancilla qubits are realized by 
\begin{gather}
\mathrm{R}^X_a(\pi) = -i (\ket{0}\bra{1}_a + \ket{1}\bra{0}_a ) \\ 
\mathrm{R}^X_q(\pi) =\ket{2}\bra{2}_q -i (\ket{0}\bra{1}_q + \ket{1}\bra{0}_q ) 
\end{gather}
and the final unitary evolution will be given by
\begin{equation}
    U = \mathrm{R}^X_a(\pi) \mathrm{R}^X_q(\pi)\mathrm{MS}^X(\pi)\mathrm{R_\text{loss}}(\phi) = U^{(0)} \otimes \mathds{1}_a + U^{(1)} \otimes X_a
\end{equation}
where
\begin{align}
    U_q^{(0)} & =\ket{1}\bra{1}_q + \cos\frac{\phi}{2} \ket{0}\bra{0}_q + \sin\frac{\phi}{2} \ket{0}\bra{2}_q,   \\
    U_q^{(1)} & = \sin\frac{\phi}{2} \ket{2}\bra{0}_q - \cos\frac{\phi}{2} \ket{2}\bra{2}_q.
\end{align}
If we assume that no population is present initially in the $\ket{2}_q$ state the operators $U^{(0)}$ and $U^{(1)}$ will reduce to
\begin{align}
    U_q^{(0)} & =\ket{1}\bra{1}_q + \cos\frac{\phi}{2} \ket{0}\bra{0}_q, \\
    U_q^{(1)} & = \sin\frac{\phi}{2} \ket{2}\bra{0}_q.
\end{align}
The single qubit process arising from the QND measurement and acting on the code qubit $q$ can be then described by two maps $\mathcal{E}_0$ and $\mathcal{E}_1$ defined as follows
\begin{gather}
    \mathcal{E}_0\colon \rho \mapsto U_q^{(0)}\rho U_q^{(0)\dag} \\
    \mathcal{E}_1\colon \rho \mapsto U_q^{(1)}\rho_q U_q^{(1)\dag} 
\end{gather}
and effectively acting on the system of code qubits as 
\begin{equation}
    \rho \mapsto \mathcal{E}_0(\rho) \otimes \ket{0}\bra{0}_a +  \mathcal{E}_1(\rho) \otimes \ket{1}\bra{1}_a.
\end{equation}
This single-qubit dynamics can be also described in the Choi representation \cite{Nielsen:2011:QCQ:1972505} by the following single qubit Choi matrices in the elementary basis $\{\ket{00},\ket{01},\ket{10},\ket{11}\}$
\begin{equation}
    \Phi^{(0)} = \frac{1}{2}\begin{pmatrix}
    \cos^2\frac{\phi}{2}  & 0 & 0  &  \cos\frac{\phi}{2} \\
    0                       & 0 & 0  & 0 \\ 
    0                       & 0 & 0  & 0 \\
    \cos\frac{\phi}{2}  & 0& 0  &  1\\    
    \end{pmatrix}, \quad \quad
    \Phi^{(1)} = \frac{1}{2}\begin{pmatrix}
    0  & 0 & 0  & 0 \\
    0  & 0 & 0  & 0 \\ 
    0  & 0 & \sin^2\frac{\phi}{2}  & 0 \\
    0 & 0& 0    & 0\\    
    \end{pmatrix}.
    \label{eqn_choi-matrix}
\end{equation}

If we now consider the effects of the controlled loss and subsequent QND loss detection on the logical states we will have that after the measurement of the ancilla the logical states $\ket{0_L} $,  the $\ket{1_L} $ and the $\ket{+i_L}  =  (\ket{0_L} + i \ket{1_L})/\sqrt{2} $ will become
\begin{gather}
\ket{0_L} \otimes \ket{0_a} \mapsto \begin{dcases}
 \ket{2000} & \text{with probability  } p_\text{L} =  \frac{1}{2} \sin^2{\frac{\phi}{2}}\\
\frac{ \cos{\frac{\phi}{2}} \ket{0000} +\ket{1111}}{(1 + \cos^2(\phi/2))^{1/2}}  & \text{with probability } 1 - p_\text{L} 
\end{dcases} \\
\ket{1_L} \otimes \ket{0_a} \mapsto \begin{dcases}
 \ket{2001} & \text{with probability  } p_\text{L} =  \frac{1}{2} \sin^2{\frac{\phi}{2}}\\
\frac{ \cos{\frac{\phi}{2}} \ket{0001} +\ket{1110}}{(1 + \cos^2(\phi/2))^{1/2}}  & \text{with probability } 1 - p_\text{L} 
\end{dcases}\\
\ket{+i_L} \otimes \ket{0_a} \mapsto \begin{dcases}
 \ket{2000} +  i \ket{2001} & \text{with probability  } p_\text{L} =  \frac{1}{4} \sin^2{\frac{\phi}{2}}\\
\frac{ \cos{\frac{\phi}{2}}\left( \ket{0000} + i \ket{0001} \right) +  \ket{1111} + i \ket{1110}}{(2 + 2\cos^2(\phi/2))^{1/2}}  & \text{with probability } 1 - p_\text{L} \label{eqn_noloss}
\end{dcases}
\end{gather}
Note that for example for the four data qubits initially prepared in the $\ket{+i_L}$ state, and if the ancilla qubit is found in the QND detection in state $\ket{0}_a$ (i.e.~no loss detected), this non-unitary time evolution results in the following (ideal) expectation value of the $X$-type stabilizer 
\begin{equation}
\braket{S^X_1} = \braket{X_1X_2X_3X_4} = \frac{4 \cos \left({\phi }/{2}\right)}{3 + \cos \phi} \approx 1-\frac{\phi^4}{128},
\label{eqn_expt_value}
\end{equation}
where the approximation in the last step holds for small loss rates, i.e.~$\phi \ll 1$.

In the following, complementary experimental data characterizing the QND loss detection unit depicted in Fig.~\ref{fig_DetectionResults} in the main text is presented.

\begin{figure*}[h!]
    \centering
    \includegraphics[width=12cm]{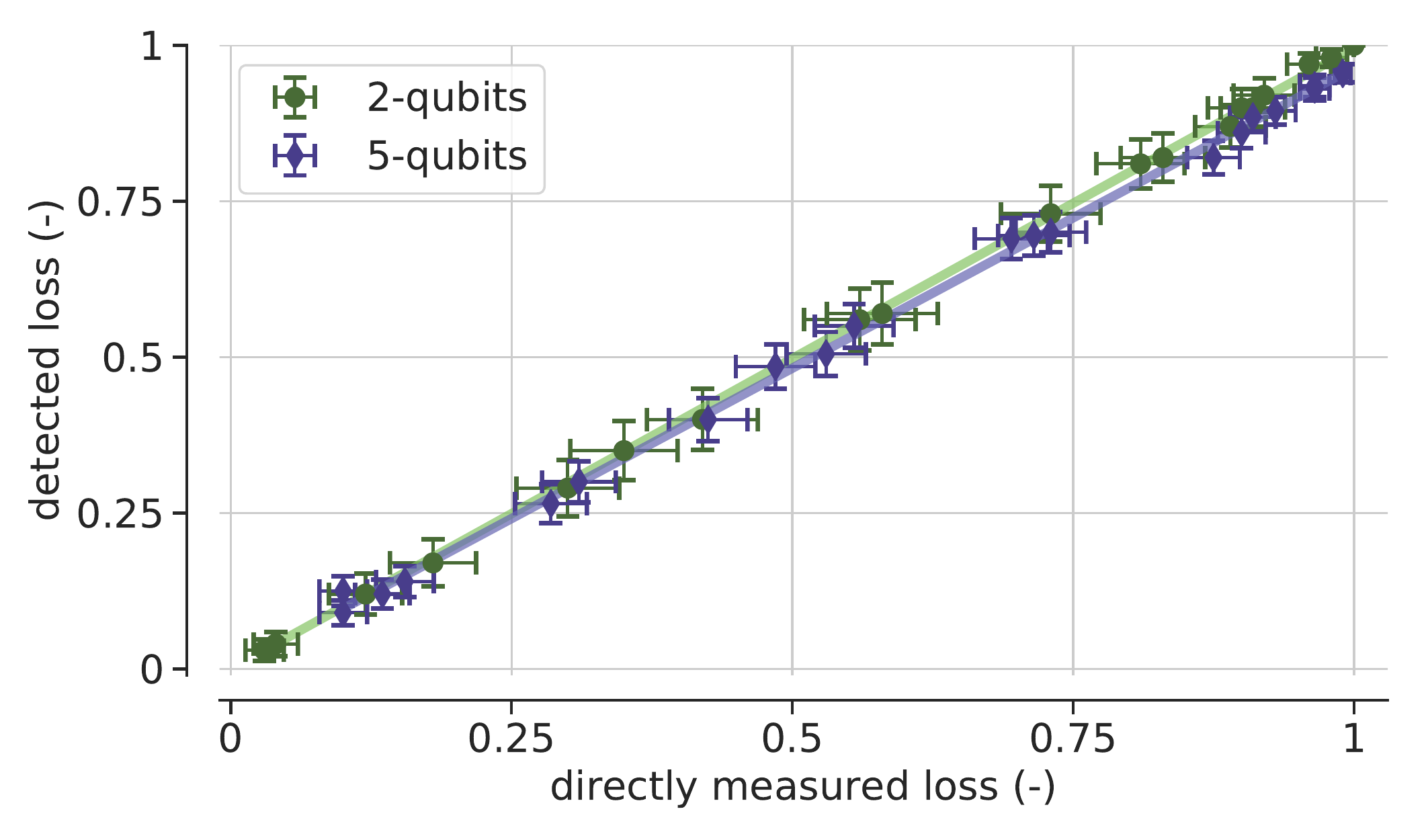}
    \caption{\textbf{Investigating the performance of the 2-qubit QND loss detection unit from Fig.~\ref{fig_KitaevCircuit}B.} In addition to the results on the full 5-qubit string depicted in Fig.~\ref{fig_DetectionResults}B we compare to an isolated experiment on 2-qubits only. Population in the $D_{5/2}$-state of qubit 1 (directly measured loss) and ancilla qubit (detected loss) measured after loss detection. Controlled loss up to \SI{100}{\%} with respect to $\ket{0}$ was introduced. The estimated detection efficiencies for the 5-qubit and 2-qubit system are \SI{96.5(4)}{\%} and \SI{99.6(3)}{\%}, respectively. Error bars correspond to 1 standard deviation of statistical uncertainty due to quantum projection noise. This demonstrates that the occurrence of a loss event can be reliably mapped onto the ancilla qubit and read out in a QND fashion.}
    \label{fig_QND_Rabi_comparison}
\end{figure*}

We start by further analyzing the performance of mapping loss onto the ancilla qubit. Results presented so far in Fig.~\ref{fig_DetectionResults} in the main text were performed on the full 5-qubit string according to the loss detection and correction circuit. To study the effect of the extra three qubits, i.e. the effect of imperfect hiding and unhiding operations, we repeat the experiments isolated on a 2-qubit string. The loss detection sub-circuit is tested by driving the loss transition $\mathrm{R_\text{loss}}(\phi)$ on  qubit 1 and measuring the population in the $D_{5/2}$-state on both qubit 1 and ancilla qubit. This measurement does not distinguish between the different Zeeman sublevels of the $D_{5/2}$-state manifold. Fig~\ref{fig_QND_Rabi_comparison} shows that in both cases loss detected by the ancilla qubit matches the loss induced on qubit 1 within statistical uncertainties. The quantified detection efficiency for the full 5-qubit string is \SI{96.5(4)}{\%}, with a false positive rate of 3$(\substack{+1 \\ -1})$~\% and a false negative rate of 1$(\substack{+1 \\ -1})$~\%. In the 2-qubit case the detection efficiency is \SI{99.6(3)}{\%} with a false positive rate of 0.6$(\substack{+1 \\ -0.6})$~\% and a false negative rate of 0.2$(\substack{+0.1 \\ -0.1})$~\%. The difference in detection efficiency is mainly due to imperfect hiding and unhiding operations induced by single-qubit addressing errors. The error bars in Fig.~\ref{fig_DetectionResults}B and ~\ref{fig_QND_Rabi_comparison} correspond to 1 standard deviation of statistical uncertainty due to quantum projection noise. For the 5-qubit (2-qubit) case 200 (100) experimental cycles were implemented.

\begin{figure*}[h!]
    \centering
    \includegraphics[width=14cm]{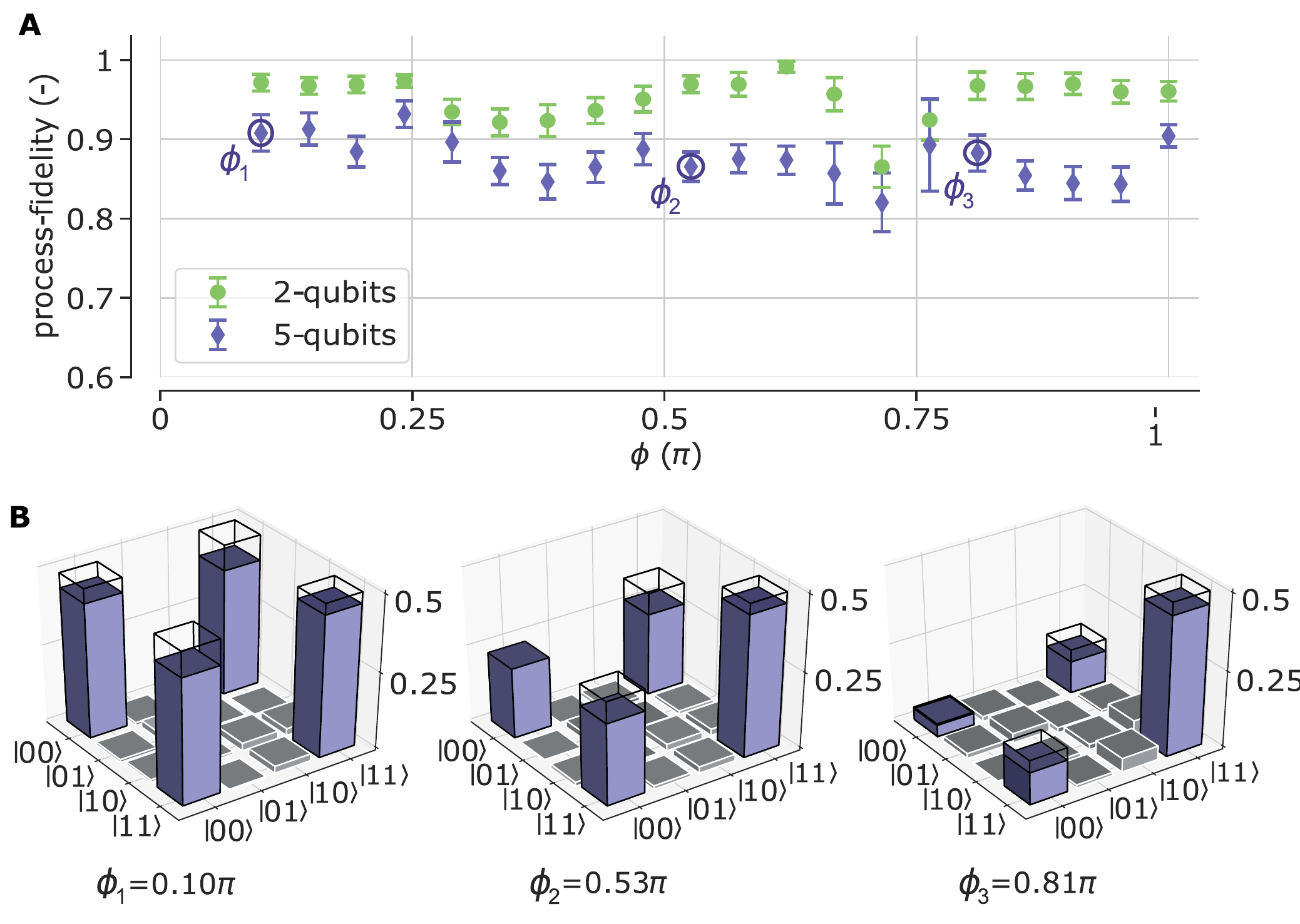}
    \caption{\textbf{Tomographic reconstruction of the non-unitary single qubit map $\Phi^{(0)}$ of Eq.~\ref{eqn_choi-matrix}, characterizing the QND measurement in the no loss case}. (\textbf{A}) Process-fidelities of the non-unitary map $\Phi^{(0)}$, when working on both the full 5-qubit string and isolated on 2-qubits only. Error bars correspond to 1 standard deviation of statistical uncertainty due to quantum projection noise. (\textbf{B}) Single qubit choi matrices $\Phi^{(0)}$ in the elementary basis $\{\ket{00},\ldots,\ket{11}\}$, reconstructed from the full 5-qubit string for loss rates indicated by the circles in Fig.~A above. The Ideal Choi-operators, according to the map $\Phi^{(0)}$ of Eq.~\ref{eqn_choi-matrix}, are denoted by the underlying black frames.}
    \label{fig_process}
\end{figure*}

Next, we present our experimental findings on the single qubit process describing the QND detection according to Eq.~\ref{eqn_choi-matrix}. We explicitly focus on the non-unitary map $\Phi^{(0)}$ characterizing the no loss case. Therefore generalized single qubit quantum process tomography was applied to qubit 1, whereupon the single qubit Choi matrices were reconstructed in the elementary basis $\{\ket{00},...,\ket{11}\}$. Experiments were implemented on both the full 5-qubit string as well as isolated on a 2-qubit string. The estimated process fidelities with the ideal non-unitary map $\Phi^{(0)}$ are shown in Fig.~\ref{fig_process}A together with plots of the associated reconstructed single qubit Choi matrices for loss rates $\phi \in \lbrace 0.10\pi, 0.53\pi, 0.81\pi\rbrace$ in Fig.~\ref{fig_process}B. In order to estimate the uncertainty of the values presented here and in Fig.~\ref{fig_DetectionResults}C we re-sample the data given by generalized quantum process tomography via a multinomial distribution and assigned the respective standard deviation, received from 100 iterations, as the statistical uncertainty. 

\begin{figure*}[h!]
    \centering
    \includegraphics[width=12cm]{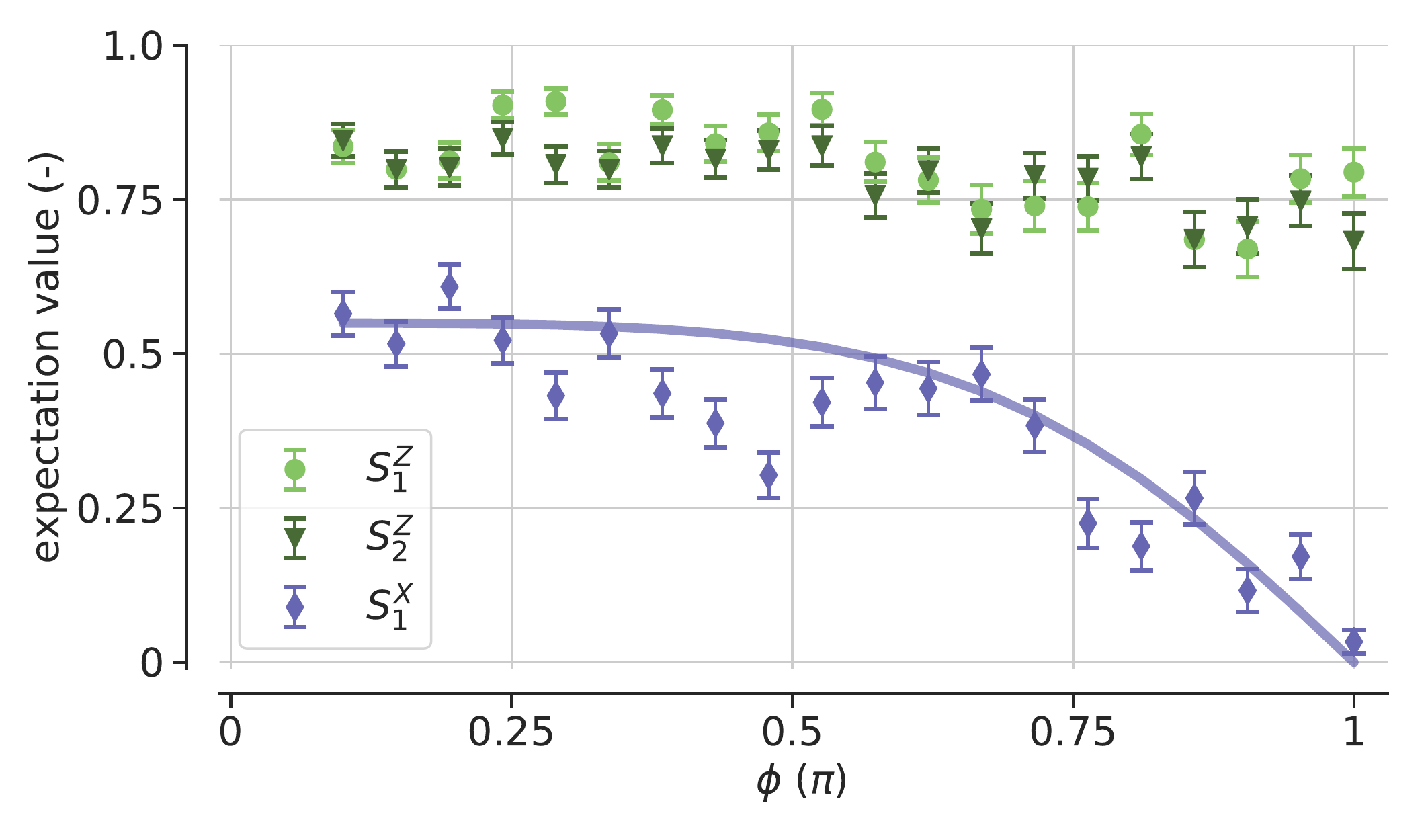}
    \caption{\textbf{Experimental investigations on the effect of the controlled loss on the logical state $\ket{+i_L}$ after the ancilla qubit measurement.} We explicitly study the no loss case. The maximum expectation values for the Z-stabilizers remain unaffected by the loss, whereas the expectation value for $S_1^{X}$ drops with increasing loss rate according to Eqs.~\ref{eqn_noloss} and \ref{eqn_expt_value}. The simulated curve represents the ideal outcome biased with the experimental measured $S^X_1$, extracted from the lowest loss rate at $\phi = 0.1\pi$. The error bars correspond to 1 standard deviation of statistical uncertainty due to quantum projection noise.}
    \label{fig_StabilizerLossratio}
\end{figure*}

In the final paragraph of this section we investigate the effect of the controlled loss on the logical state after the ancilla measurement. To follow this idea, we initialize $\ket{+i_L}$ and proceed with the loss detection as shown in Fig.~\ref{fig_KitaevCircuit}. Controlled loss between $\phi=0.1\pi$ and $\pi$ is introduced on qubit 1. We study the case where we find the ancilla qubit in $\ket{0}$, i.e. in the absence of loss. Fig.~\ref{fig_StabilizerLossratio} shows the results on the expectation values of the stabilizer generators $S_1^{X}$, $S_1^{Z}$ and $S_2^{Z}$. The maximum expectation values of the Z-stabilizers remain unaffected by the loss, whereas the expectation value for $S_1^{X}$ drops for increasing loss rates $\phi$ according to Eqs.~\ref{eqn_noloss} and \ref{eqn_expt_value}. The underlying modelled curve for $S_1^{X}$ represents the ideal outcome biased with the experimentally measured $S_1^{X}$ value, extracted from the lowest loss rate at $\phi = 0.1\pi$. The error bars correspond to 1 standard deviation of statistical uncertainty due to quantum projection noise. In total 200 experimental cycles were implemented. The results show that the experiment and theory predictions of the effect of loss and QND detection are in good agreement.

\section{Additional Data and experimental information on the 1+4-qubit loss detection and correction algorithm}
\label{supp:additional-data}
Here, we provide complementary results on the full 1+4-qubit loss detection and correction algorithm for the logical input states $\lbrace\ket{0_L}, \ket{1_L}, \ket{+i_L}=(\ket{0_L}+i\ket{1_L})/\sqrt(2)\rbrace$ under three different loss rates $\phi \in \lbrace 0.1\pi, 0.2\pi, 0.5\pi \rbrace$. 
Next to fidelities, expectation values for stabilizers and logical operators the remaining population in the code space $P_\text{CS}$ was estimated according to $\hat{P}_\text{CS}\Ket{\psi}=P_\text{CS}\Ket{\psi}$. Here, $\hat{P}_\text{CS}$ represents the projector onto the code space, defined as the simultaneous +1 eigenspace given by all generators of the stabilizer group $\{S_1^X, S_1^Z, S_2^Z\}$ and $\{\widetilde{S}_1^X, \widetilde{S}_1^Z\}$ for the 4-qubit and the 3-qubit logical encoding, respectively. The code space projector reads:
\begin{equation}
\hat{P}_\text{CS}=\prod_i \frac{1}{2}(1+S_x^{(i)})\prod_j\frac{1}{2}(1+S_z^{(j)}) 
\end{equation}{}
All results presented in Fig.~\ref{fig_KitaevResults} and in Tabs.~\ref{tab:addtionaldata0L}, \ref{tab:addtionaldata1L} and \ref{tab:addtionaldataiL} were extracted from full 4-qubit quantum state tomography using linear state reconstruction technique. In order to estimate the uncertainty of these values we re-sample the data given by quantum state tomography via a multinomial distribution and assigned the respective standard deviation, received from 100 iterations, as the statistical uncertainty. In order to receive enough experimental data under both loss cases for state reconstruction, we adjusted the number of experimental cycles depending on the induced loss rate. The corresponding values on cycle numbers read: 1000 cycles for $\phi = \SI{0.1}{\pi}$, 600 cycles for $\phi = \SI{0.2}{\pi}$ and 200 cycles for $\phi = \SI{0.5}{\pi}$. 
Tabs.~\ref{tab:addtionaldata0L}, \ref{tab:addtionaldata1L} and \ref{tab:addtionaldataiL} contain the entire data gained on the 1+4-qubit loss detection and correction algorithm. Each table is assigned to one of the logical input states $\lbrace\ket{0_L}, \ket{1_L}, \ket{+i_L}=(\ket{0_L}+i\ket{1_L})/\sqrt(2)\rbrace$ and includes data on in total three different loss rates $\phi \in \lbrace 0.1\pi, 0.2\pi, 0.5\pi \rbrace$.

\setlength{\tabcolsep}{0.6em}
\begin{table*}[h!]
\centering
\begin{tabular}{c c c c c c c c}
    \toprule
    \midrule
    \multirow{3}[4]{*}{} & \multicolumn{7}{c}{encoding}\\ 
    \cmidrule(rl){2-8}
    & $P_\text{CS} $  & $S_1^X$ & $S_1^Z$ & $S_2^Z$ & $T^X$ & $T^Y$ & $T^Z$ \\ 
    \cmidrule(l){2-8}
    \multicolumn{1}{r}{} & 0.93(2) & 0.84(6) & 0.95(1) & 0.94(1) & 0.01(2) & -0.01(3) & 0.93(1)
    \vspace{0.25cm}\\
    \multirow{2}[4]{*}{} & \multicolumn{7}{c}{no-loss}\\ 
    \cmidrule(rl){1-1}\cmidrule(rl){2-8}
    $\phi$ ($\pi$) & $P_\text{CS} $  & $S_1^X$ & $S_1^Z$ & $S_2^Z$ & $T^X$ & $T^Y$ & $T^Z$ \\ 
    \cmidrule(rl){1-1}\cmidrule(l){2-8}
    \multicolumn{1}{r}{\SI{0.1}{}}  & 0.74(1) & 0.60(2) & 0.86(1) & 0.85(1) & -0.02(1) & -0.13(1)& 0.84(1)   \\
    \multicolumn{1}{r}{\SI{0.2}{}} & 0.72(1) & 0.57(4) & 0.87(1) & 0.87(1) & -0.02(1) & -0.14(1) & 0.84(1)   \\
    \multicolumn{1}{r}{\SI{0.5}{}} & 0.68(2) & 0.40(8) & 0.80(2) & 0.82(2) & -0.06(2) & -0.10(3) & 0.79(2)      
    \vspace{0.25cm} \\
    \multirow{2}[4]{*}{} & \multicolumn{7}{c}{loss}\\ 
    \cmidrule(rl){1-1}\cmidrule(rl){2-8}
    $\phi$ ($\pi$) & $P_\text{CS} $  & $\widetilde{S}_1^X$ & $\widetilde{S}_1^Z$ &  & $\widetilde{T}^X$ & $\widetilde{T}^Y$ & $\widetilde{T}^Z$ \\ 
    \cmidrule(rl){1-1}\cmidrule(l){2-8}
    \multicolumn{1}{r}{\SI{0.1}{}} & 0.44(4) & 0.19(16) & 0.60(5) & & 0.00(3) & 0.06(5) & 0.53(4)   \\
    \multicolumn{1}{r}{\SI{0.2}{}} & 0.65(5) & 0.53(15) & 0.51(5) & & 0.00(3) & -0.06(6) & 0.63(4)   \\
    \multicolumn{1}{r}{\SI{0.5}{}} & 0.69(4) & 0.63(11) & 0.59(4) & & 0.00(2) & -0.05(4) & 0.65(3)   \\
    \midrule
    \bottomrule
\end{tabular}
\caption{\textbf{logical state $\ket{0_L}$}: Complementary experimental data on the 1+4-qubit loss detection and correction algorithm (see Fig.~\ref{fig_KitaevResults} in the main text) including results on three different loss rates $\phi$.}
\label{tab:addtionaldata0L}
\end{table*}

\begin{table*}[h!]
\centering
\begin{tabular}{c c c c c c c c}
    \toprule
    \midrule
    \multirow{3}[4]{*}{} & \multicolumn{7}{c}{encoding}\\ 
    \cmidrule(rl){2-8}
    & $P_\text{CS} $  & $S_1^X$ & $S_1^Z$ & $S_2^Z$ & $T^X$ & $T^Y$ & $T^Z$ \\ 
    \cmidrule(l){2-8}
    \multicolumn{1}{r}{} & 0.91(1) & 0.74(8) & 0.94(1) & 0.95(1) & -0.01(2) & 0.04(3) & -0.93(1)  
    \vspace{0.25cm} \\
    \multirow{2}[4]{*}{} & \multicolumn{7}{c}{no-loss}\\ 
    \cmidrule(rl){1-1}\cmidrule(rl){2-8}
    $\phi$ ($\pi$) & $P_\text{CS} $  & $S_1^X$ & $S_1^Z$ & $S_2^Z$ & $T^X$ & $T^Y$ & $T^Z$ \\ 
    \cmidrule(rl){1-1}\cmidrule(l){2-8}
    \multicolumn{1}{r}{\SI{0.1}{}}  & 0.79(1) & 0.68(3) & 0.86(1) & 0.85(1) & -0.02(1) & 0.10(1) & -0.83(1)   \\
    \multicolumn{1}{r}{\SI{0.2}{}} & 0.78(1) & 0.67(3) & 0.86(1) & 0.86(1) & -0.03(1) & 0.10(2) & -0.81(1)   \\
    \multicolumn{1}{r}{\SI{0.5}{}} & 0.72(1) & 0.61(6) & 0.78(2) & 0.80(2) & 0.04(1) & 0.07(3) & -0.74(2)      \vspace{0.25cm} \\
    \multirow{2}[4]{*}{} & \multicolumn{7}{c}{loss}\\ 
    \cmidrule(rl){1-1}\cmidrule(rl){2-8}
    $\phi$ ($\pi$) & $P_\text{CS} $  & $\widetilde{S}_1^X$ & $\widetilde{S}_1^Z$ &  & $\widetilde{T}^X$ & $\widetilde{T}^Y$ & $\widetilde{T}^Z$ \\ 
    \cmidrule(rl){1-1}\cmidrule(l){2-8}
    \multicolumn{1}{r}{\SI{0.1}{}} & 0.44(6) & 0.23(18) & 0.44(5) & & 0.00(3) & -0.04(5) & -0.43(6)   \\
    \multicolumn{1}{r}{\SI{0.2}{}} & 0.63(5) & 0.49(14) & 0.61(4) & & -0.02(3) & 0.01(5) & -0.59(5)   \\
    \multicolumn{1}{r}{\SI{0.5}{}} & 0.80(3) & 0.80(8) & 0.80(3) & & -0.09(3) & 0.09(5) & -0.72(3)   \\
    \midrule
    \bottomrule
\end{tabular}
\caption{\textbf{logical state $\ket{1_L}$}: Complementary experimental data on the 1+4-qubit loss detection and correction algorithm (see Fig.~\ref{fig_KitaevResults} in the main text) including results on three different loss rates $\phi$.}
\label{tab:addtionaldata1L}
\end{table*}

\begin{table*}[h!]
\centering
\begin{tabular}{c c c c c c c c}
    \toprule
    \midrule
    \multirow{3}[4]{*}{} & \multicolumn{7}{c}{encoding}\\ 
    \cmidrule(rl){2-8}
    & $P_\text{CS} $  & $S_1^X$ & $S_1^Z$ & $S_2^Z$ & $T^X$ & $T^Y$ & $T^Z$ \\ 
    \cmidrule(l){2-8}
    \multicolumn{1}{r}{} & 0.88(2) & 0.61(8) & 0.95(1) & 0.97(1) & 0.02(2) & 0.96(1) & 0.04(4)  
    \vspace{0.25cm} \\
    \multirow{2}[4]{*}{} & \multicolumn{7}{c}{no-loss}\\ 
    \cmidrule(rl){1-1}\cmidrule(rl){2-8}
    $\phi$ ($\pi$) & $P_\text{CS} $  & $S_1^X$ & $S_1^Z$ & $S_2^Z$ & $T^X$ & $T^Y$ & $T^Z$ \\ 
    \cmidrule(rl){1-1}\cmidrule(l){2-8}
    \multicolumn{1}{r}{\SI{0.1}{}}  & 0.77(1) & 0.60(2) & 0.88(1) & 0.87(1) & -0.02(1) & 0.81(1) & 0.00(1)   \\
    \multicolumn{1}{r}{\SI{0.2}{}} & 0.73(1) & 0.58(4) & 0.88(1) & 0.85(1) & -0.03(1) & 0.79(1) & 0.01(1)   \\
    \multicolumn{1}{r}{\SI{0.5}{}} & 0.67(1) & 0.48(3) & 0.82(1) & 0.82(1) & -0.07(1) & 0.75(1) & 0.02(1)      \vspace{0.25cm} \\
    \multirow{2}[4]{*}{} & \multicolumn{7}{c}{loss}\\ 
    \cmidrule(rl){1-1}\cmidrule(rl){2-8}
    $\phi$ ($\pi$) & $P_\text{CS} $  & $\widetilde{S}_1^X$ & $\widetilde{S}_1^Z$ &  & $\widetilde{T}^X$ & $\widetilde{T}^Y$ & $\widetilde{T}^Z$ \\ 
    \cmidrule(rl){1-1}\cmidrule(l){2-8}
    \multicolumn{1}{r}{\SI{0.1}{}} & 0.52(6) & 0.25(16) & 0.57(5) & & 0.11(3) & 0.51(5) & -0.01(7)   \\
    \multicolumn{1}{r}{\SI{0.2}{}} & 0.61(5) & 0.55(14) & 0.56(4) & & 0.13(3) & 0.47(4) & 0.06(5)   \\
    \multicolumn{1}{r}{\SI{0.5}{}} & 0.80(2) & 0.81(4) & 0.80(2) & & 0.13(1) & 0.80(2) & 0.01(2)   \\
    \midrule
    \bottomrule
\end{tabular}
\caption{\textbf{logical state $\ket{+i_L}$}: Complementary experimental data on the 1+4-qubit loss detection and correction algorithm (see Fig.~\ref{fig_KitaevResults} in the main text) including results on three different loss rates $\phi$.}
\label{tab:addtionaldataiL}
\end{table*}

\section{Imperfections in the QND loss detection}
\label{supp:imperfections-QND}
From the data in the previous section in the case of low loss rates, namely $\phi \in\lbrace 0.1\pi, 0.2\pi\rbrace$, we find that the probability of success for reconstructing the code after loss is lower than for the higher loss rate $\phi = 0.5\pi$. We relate this to imperfections in the QND loss detection unit. Let's assume the error on the detection unit is of the same order as the loss rate, then many of the measurement cycles detected as loss will be false positives. This limits the performance for current system parameters in the regime of low qubit loss rates.

In order to quantitatively study this effect, we model imperfections in the QND loss detection by a depolarizing noise-channel on each individual qubit. Since loss is induced on the lower qubit state $S_{1/2}(m=-1/2)=\ket{0}$, complete loss ($\phi=\pi$) of this state leads to an overall loss rate of 50\% for the encoded GHZ-state, where half of the population occupies the $D_{5/2}(m=-1/2)=\ket{1}$ state. Taking this into account (with a factor $0.5$ in front of the $\sin^2$-term) our model reads:
\begin{align}
\rho \longrightarrow \frac{p}{3} \sum_{k=1}^3 M_{(k)}(\rho) + (1-p)\rho \hspace{0.75cm} \text{with}\hspace{0.4cm} & M_{(k)}(\rho) = \frac{1}{4}\sum_{i \in \lbrace x, y, z, id\rbrace} \sigma^{(k)\dagger}_i\rho\sigma^{(k)}_i \hspace{1cm}\\
\text{and}\hspace{0.4cm}\ &p = \frac{p_\text{QND}}{p_\text{QND}+0.5\sin^2(\phi/2)}.
\end{align}

In Fig.~\textcolor{blue}{S5} we plot the model against the measured data prepared in logical $\ket{1_L}$, previously presented in Tab.~\ref{tab:addtionaldata1L}. We find good agreement between model and data for $p_\text{QND} = \SI{3.3}{\%}$.

\begin{figure*}[h!]
    \centering
    \includegraphics[width=9.5cm]{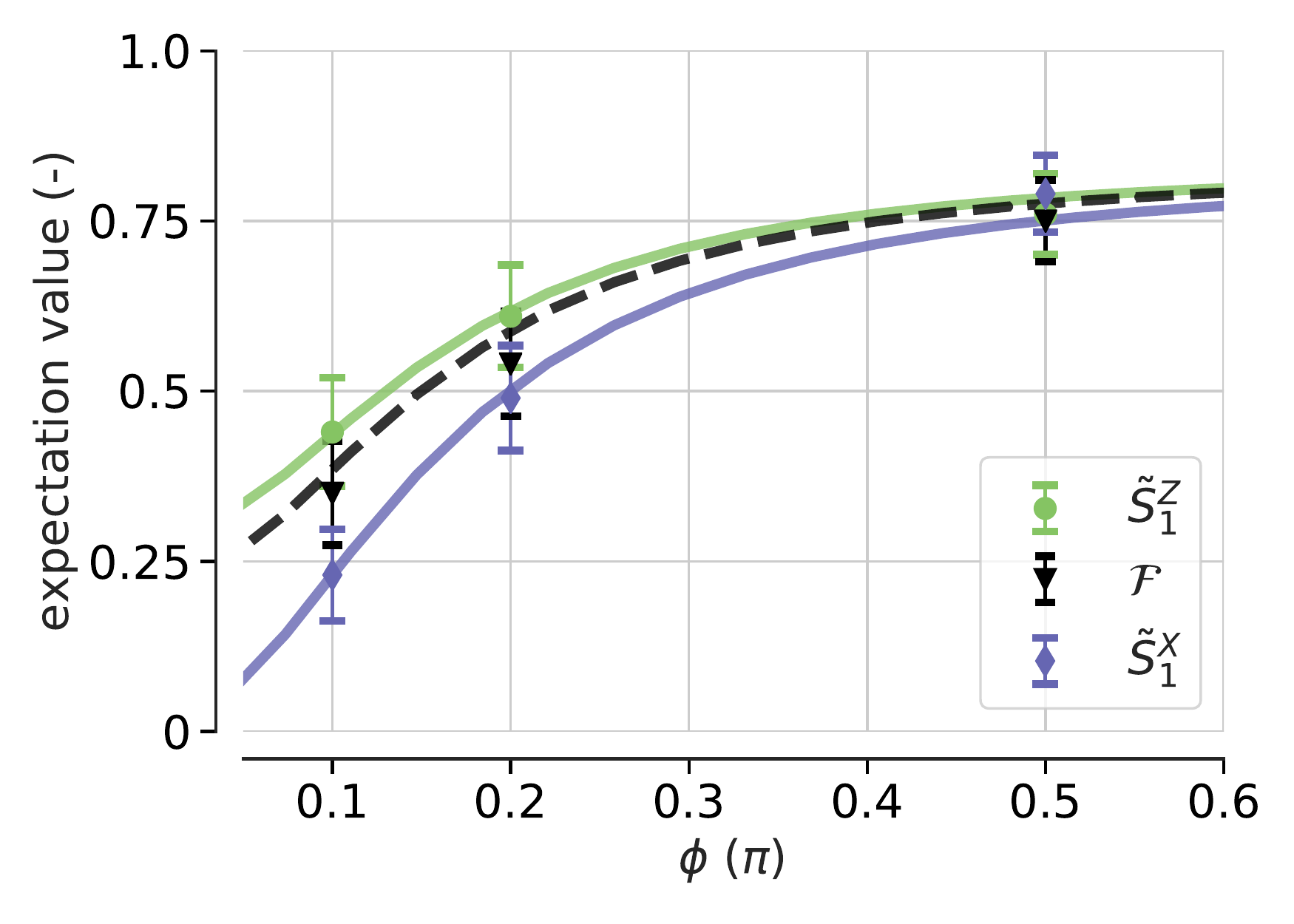}
    \caption{\textbf{Modelling an imperfect QND-detection scheme under the assumption of an independent depolarizing noise channel on each qubit.} The theoretical model (lines) shows good agreement with the experimental data (points with error bars). The error bars correspond to 1 standard deviation of statistical uncertainty due quantum projection noise.}
    \label{fig_losssimulation}
\end{figure*}

The imperfections originate mainly from addressing errors when hiding and unhiding the qubits 2, 3 and 4, preventing them from taking part in the QND detection unit. For faulty experimental shots, where one of those qubits is not hidden in the upper $D_{5/2}(m=+1/2)$ level, it will affect the loss detection measurement. Hence it is likely to happen that the particular experimental cycle assigns to the wrong loss case. By improving the addressing optics such errors could be further suppressed.

\end{document}